\documentclass[sigconf]{acmart}
\usepackage{multirow}
\usepackage{enumitem}
\usepackage{arydshln}
\usepackage{graphicx} 		
\usepackage{subfigure}		

\AtBeginDocument{%
  \providecommand\BibTeX{{%
    \normalfont B\kern-0.5em{\scshape i\kern-0.25em b}\kern-0.8em\TeX}}}

\setcopyright{acmlicensed}
\copyrightyear{2018}
\acmYear{2018}
\acmDOI{XXXXXXX.XXXXXXX}

\acmConference[Conference acronym 'XX]{Make sure to enter the correct
  conference title from your rights confirmation emai}{June 03--05,
  2018}{Woodstock, NY}
\acmISBN{978-1-4503-XXXX-X/18/06}

\begin{document}

\title{QAGCF: Graph Collaborative Filtering for Q\&A Recommendation}

\author{Changshuo Zhang}
\affiliation{%
  \institution{Gaoling School of AI,\\ Renmin University of
China}
  \city{Beijing}
  \country{China}
}
\email{lyingcs@ruc.edu.cn}

\author{Teng Shi}
\affiliation{%
  \institution{Gaoling School of AI,\\ Renmin University of
China}
    \city{Beijing}
  \country{China}
  }
\email{shiteng@ruc.edu.cn}

\author{Xiao Zhang}
\authornote{Xiao Zhang is the corresponding author.}
\affiliation{%
  \institution{Gaoling School of AI,\\ Renmin University of
China}
    \city{Beijing}
  \country{China}
  }
\email{zhangx89@ruc.edu.cn}

\author{Yanping Zheng}
\affiliation{%
  \institution{Gaoling School of AI,\\ Renmin University of
China}
    \city{Beijing}
  \country{China}
  }
\email{zhengyanping@ruc.edu.cn}

\author{Ruobing Xie}
\affiliation{%
  \institution{Wechat, Tencent}
  \city{Beijing}
  \country{China}}
\email{xrbsnowing@163.com}

\author{Qi Liu}
\affiliation{%
  \institution{Wechat, Tencent}
  \city{Beijing}
  \country{China}}
\email{addisliu@tencent.com}

\author{Jun Xu}
\affiliation{%
  \institution{Gaoling School of AI,\\ Renmin University of
China}
  \city{Beijing}
  \country{China}}
\email{junxu@ruc.edu.cn}

\author{Ji-Rong Wen}
\affiliation{%
  \institution{Gaoling School of AI,\\ Renmin University of
China}
  \city{Beijing}
  \country{China}}
\email{jrwen@ruc.edu.cn}

\renewcommand{\shortauthors}{Changshuo Zhang et al.}

\begin{abstract}

Question and answer (Q\&A) platforms usually recommend question-answer pairs to meet users' knowledge acquisition needs, unlike traditional recommendations that recommend only one item.
This makes user behaviors more complex, and presents two challenges for Q\&A recommendation, including: the \emph{collaborative information entanglement}, which means user feedback is influenced by either the question or the answer; and the \emph{semantic information entanglement}, where questions are correlated with their corresponding answers, and correlations also exist among different question-answer pairs.
Traditional recommendation methods treat the question-answer pair as a whole or only consider the answer as a single item, which overlooks the two challenges and cannot effectively model user interests.
To address these challenges, we introduce \textbf{Q}uestion \& \textbf{A}nswer \textbf{G}raph \textbf{C}ollaborative \textbf{F}iltering (QAGCF), a graph neural network model that creates separate graphs for collaborative and semantic views to disentangle the information in question-answer pairs.
The collaborative view disentangles questions and answers to individually model collaborative information, while the semantic view captures the semantic information both within and between question-answer pairs. These views are further merged into a global graph to integrate the collaborative and semantic information.
Polynomial-based graph filters are used to address the high heterophily issues of the global graph. 
Additionally, contrastive learning is utilized to obtain robust embeddings during training.
Extensive experiments on industrial and public datasets demonstrate that QAGCF consistently outperforms baselines and achieves state-of-the-art results.

\end{abstract}



\maketitle

\section{Introduction}

\begin{figure}[t]
\centering
\includegraphics[width=0.95\columnwidth]{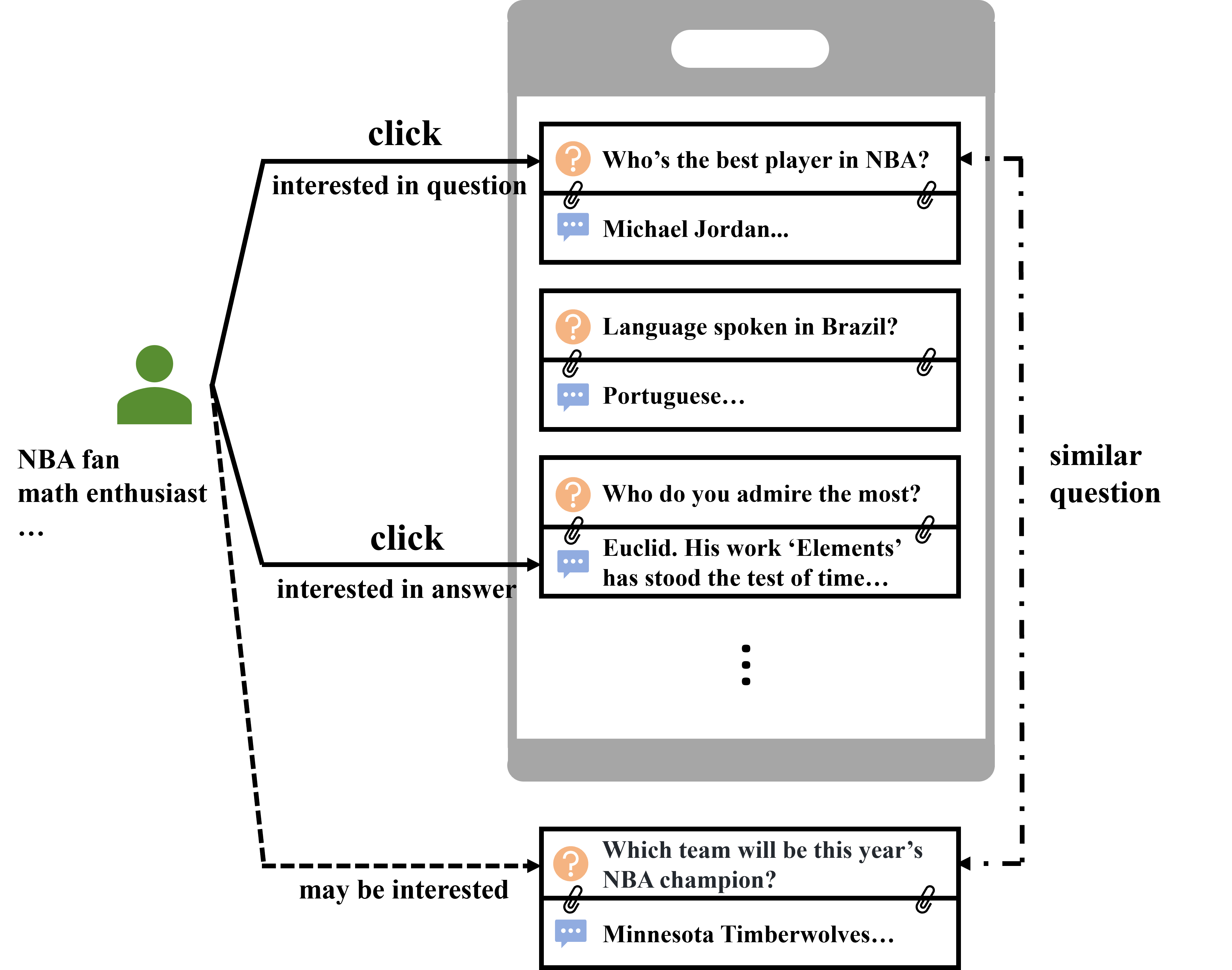}
\caption{An illustration of Q\&A Recommendation. When users access the Q\&A platform, the recommender system provides them with a list of recommended Q\&A pairs. Users can interact with Q\&A pairs they are interested in. 
}
\label{qa_rec}
\vspace{-0.5cm}
\end{figure}

Traditional recommender systems like those used in music~\cite{zhang2022counteracting, dai2024model}, e-commerce~\cite{pei2019personalized,huzhang2021aliexpress}, social media~\cite{wu2020diffnet++,wu2019dual} platforms typically focus on recommending a single item, such as a product or a video. However, in order to satisfy users' needs for acquiring and sharing knowledge, Question and Answer (Q\&A) platforms such as Yahoo Answers, Zhihu, and Quora recommend question-answer pairs to users, as shown in Figure~\ref{qa_rec}. This results in user behavior being influenced by both the question and the answer rather than just a single item.



\textbf{Motivation.} 
Existing research on Q\&A platforms has primarily focused on finding potential responders to questions~\cite{dror2011want, guo2008tapping, qu2009probabilistic}, providing high-quality answers to users' queries~\cite{jeon2006framework, hammond1997question}, or simply considering the traditional recommendation task of answers~\cite{hao2021large}.
However, research on the recommendation task on Q\&A platforms is still limited.
An intuitive approach is to treat the question-answer pair or only the answer as a single item, and then apply traditional recommendation algorithms.
However, this methodology faces two major challenges, making it an extraordinary task:

\textit{\textbf{C1:}} 
\textbf{Question-Answer collaborative information entanglement}.
In Q\&A recommendation, users are exposed to both questions and answers simultaneously, both of which can influence user behavior such as clicks. 
For example, as shown in Figure~\ref{qa_rec}, the user, being an NBA fan, clicked on the corresponding question-answer pair because of the question \textit{"\textbf{Q}: Who's the best player in NBA?"}. At the same time, the user, being a math enthusiast, clicked on the corresponding question-answer pair because of the answer \textit{"\textbf{A}: Euclid. His work 'element'..."}.
Disentangling the collaborative information from question and answer can help us better understand the reason of user clicks.
Treating the question and answer as a whole fails to disentangle the collaborative information specific to each of them, while solely focusing on the answer overlooks the collaborative information from the question.

\textit{\textbf{C2:}} 
\textbf{Question-Answer semantic information entanglement}.
On one hand, within a question-answer pair, there exists semantic similarity between the question and the answer. As shown in Figure~\ref{qa_rec}, both \textit{"\textbf{Q}: Who's the best player in NBA?"} and \textit{"\textbf{A}: Michael Jordan..."} are related to the NBA.
On the other hand, there is also semantic similarity among different question-answer pairs. As shown in Figure~\ref{qa_rec}, \textit{"\textbf{Q}: Who's the best player in NBA?"} and \textit{"\textbf{Q}: Which team will be this year's NBA champion?"} correspond to question-answer pairs that are both related to the NBA.
The similarity among different question-answer pairs can cause users to click on similar pairs after clicking on one.
Treating the question and answer as a whole or only considering the answer cannot disentangle the semantic information within the question-answer pair or among different question-answer pairs.
\begin{figure}[t]
	\centering
	\subfigure[\textit{Challenge \textbf{C1}} validation] {\includegraphics[width=.45\columnwidth]{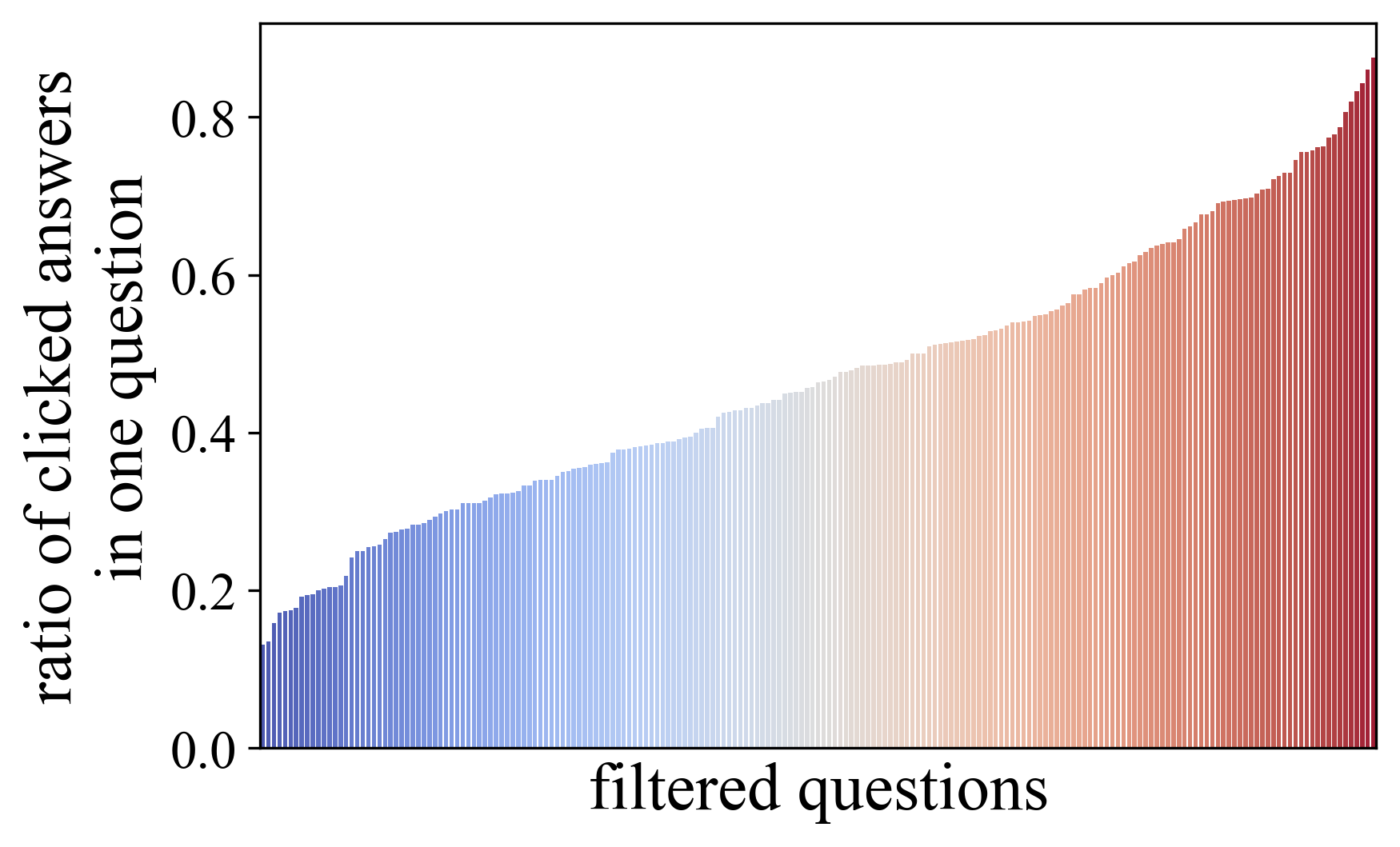}}
	\subfigure[\textit{Challenge \textbf{C2}} validation] {\includegraphics[width=.45\columnwidth]{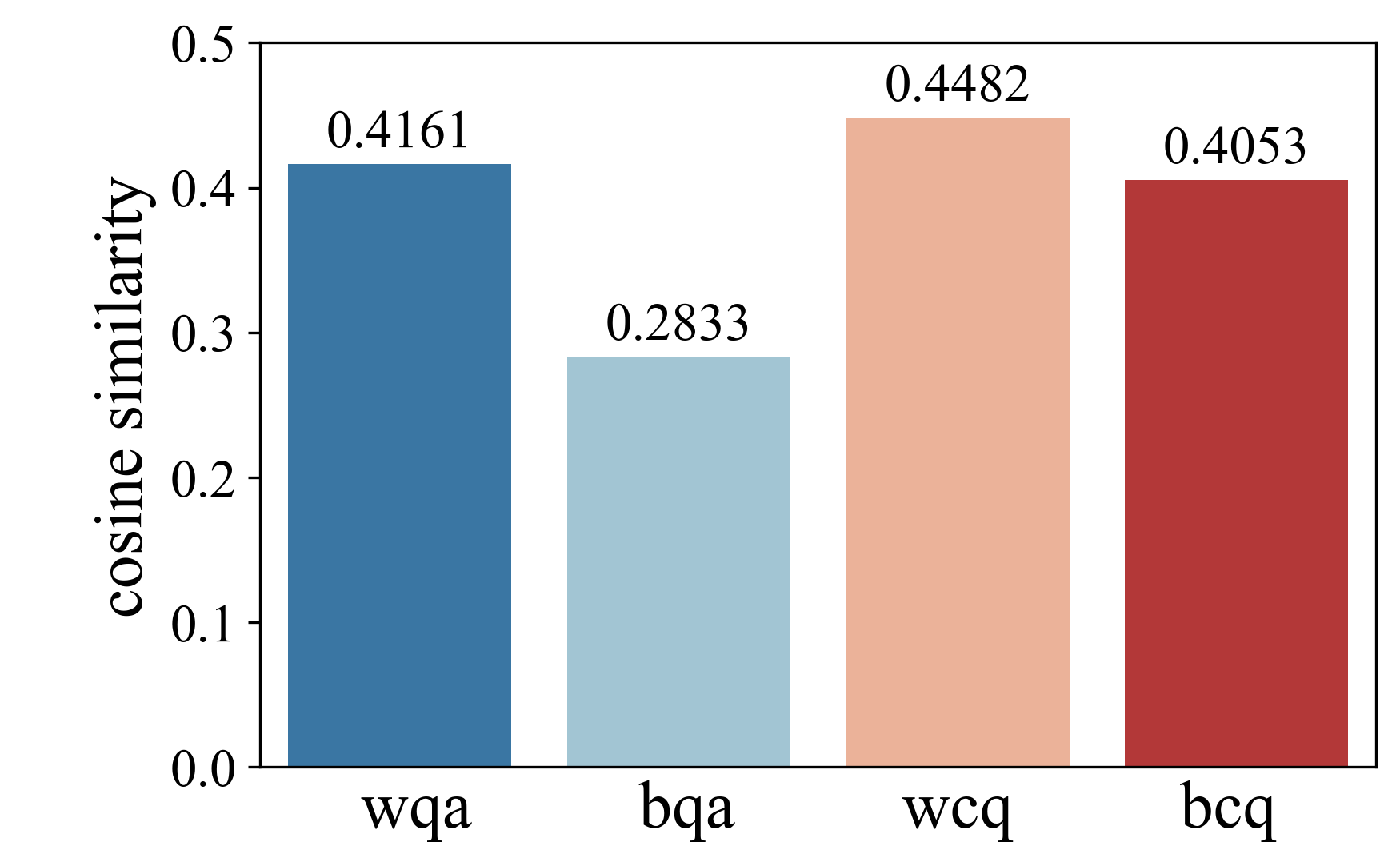}}
    \vspace{-5px}
	\caption{Analysis of the two challenges on ZhihuRec. 
    (a): We filter out questions that have been exposed more than 10 times and have more than 10 answers\protect\footnotemark[1]. 
    (b): We analyzed the similarity in four groups: wqa (within question-answer pairs), bqa (between questions and unrelated answers), weq (within clicked questions), beq (between clicked questions and unclicked ones) }
	\label{intro:ana}
    \vspace{-0.5cm}
\end{figure}
\footnotetext[1]{For those questions with fewer exposures or answers, users may click for other factors, making it difficult to determine whether the click is due to the question or the answer.}

To further validate the existence of the two challenges, we conduct an analysis based on the ZhihuRec dataset~\cite{hao2021large}.
We validate the first challenge \textit{\textbf{C1}} by calculating the ratio of the number of clicked answers to the total number of answers for each question. 
An intuitive example is, if a question is exposed multiple times and most of the answers are clicked, we can consider that the click is caused by the question; conversely, when a question is exposed multiple times but only a small proportion of the answers are clicked, we can consider that the click is due to the answers. The analysis results, as shown in Figure~\ref{intro:ana}(a), indicate that in all questions selected, this ratio covers almost the entire range from $0$ to $1.0$, suggesting that users might click due to the questions as well as the answers. To validate the second challenge \textit{\textbf{C2}}, 
We first analyze the similarity\protect\footnotemark[2]\footnotetext[2]{The similarity is calculated as the cosine similarity between embeddings of the questions/answers, where embeddings are computed as the average of all token embeddings provided in the dataset.} within question-answer pairs and between questions and unrelated answers. The results are shown on the left of Figure~\ref{intro:ana}(b), with an average similarity of $0.4161$ within question-answer pairs, and a similarity of $0.2833$ between questions and unrelated answers. This indicates that the semantic similarity between question-answer pairs is higher.
Next, we analyze the similarity within the questions clicked by each user and between the clicked and unclicked questions. The results are shown on the right of Figure~\ref{intro:ana}(b), indicating an average similarity of $0.4482$ for all user-clicked questions, and a similarity of $0.4053$ between clicked and unclicked questions. This suggests that there exists similarity between different question-answer pairs, which can lead users to click on other similar pairs after clicking on one.
Through detailed validation and analysis experiments, we confirm the existence of these two challenges.

\textbf{Contribution. }
To address the above challenges, we propose a novel recommendation model called \textbf{Q}uestion \& \textbf{A}nswer \textbf{G}raph \textbf{C}ollaborative \textbf{F}iltering (QAGCF) based on graph neural networks~\cite{wang2019neural,he2020lightgcn}. 
Firstly, we create graphs separately from collaborative and semantic views to disentangle the collaborative and semantic information of question-answer pairs.
Specifically, for the collaborative view, we disentangle the collaborative relationships between users and questions/answers, constructing two bipartite graphs: the User-Question (U-Q) graph and the User-Answer (U-A) graph. 
For the semantic view, on the one hand, we establish edges between questions and their corresponding answers, using similarity as weights. On the other hand, for different question-answer pairs, we connect questions with high similarity and construct the Question-Answer (Q-A) graph and the Question-Question (Q-Q) graph.
Then, we merge the graphs constructed from the collaborative and semantic views into a global graph to integrate the disentangled collaborative and semantic information.
This facilitates subsequent modules in extracting global embeddings for users, questions, and answers.
Due to the global graph's high heterophily~\cite{he2022convolutional, lei2022evennet}, we train the graph by using polynomial-based graph filters~\cite{guo2023manipulating}. 
Additionally, to prevent overfitting and learn more robust embeddings, we introduce contrastive learning~\cite{wu2021self, yu2022graph, yu2023xsimgcl} techniques during training.


\begin{figure*}[t]
\centering
\includegraphics[width=\textwidth]{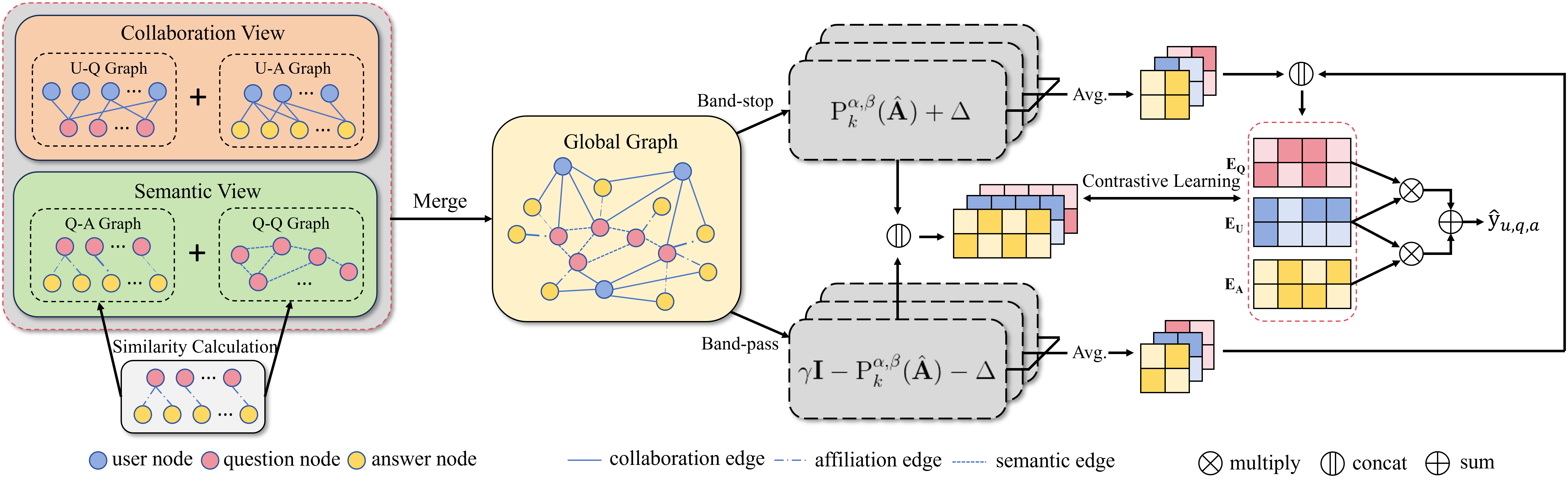}
\caption{
The overall framework of QAGCF consists of four modules: (1) Construction of Collaborative View and semantic View, (2) Band-stop and Band-pass Filters for processing information in different frequency bands of the Global Graph, (3) Contrastive Learning for noise-based embedding augmentation, (4) Prediction module based on the learned embeddings.}
\label{main_model2}
\end{figure*}

Our contribution can be summarized as follows:
\begin{itemize}[leftmargin=*]
    \item We have identified two challenges in the Q\&A recommendation task: Question-Answer collaborative and semantic information entanglement. Through experimental analysis, we have verified the existence and universality of these challenges.
    \item We propose a graph neural network model, QAGCF, for Q\&A recommendation task. By constructing collaborative and semantic views, we disentangle the collaborative and semantic information of question-answer pairs.
    Polynomial-based graph filters are further used for the global graph which integrates the collaborative and semantic information.
    \item We conducted extensive experiments on two real-world Q\&A recommendation datasets to demonstrate the superiority of the proposed model. Further ablation studies explain the superiority of our designed modules.
\end{itemize}

\section{Problem Formulation}
We present the problem formulation for the Q\&A recommendation task. Given a set of users $\mathcal{U}=\left\{u_1, u_2, \cdots, u_M\right\}$, a set of questions $\mathcal{Q}=$ $\left\{q_1, q_2, \cdots, q_N\right\}$, and a set of answers $\mathcal{A}=\left\{a_1, a_2, \cdots, a_O\right\}$, where $M$, $N$, and $O$ are the number of users, questions, and answers, respectively. 
The user-question interactions, user-answer interactions, and question-answer affiliations are denoted as
$\mathbf{X} \in \{0,1\}^{M\times N}$, $\mathbf{Y} \in \{0,1\}^{M\times O}$, and $\mathbf{Z} \in \{0,1\}^{N\times O}$, where the entry of each matrix $X_{u q}=1, Y_{u a}=1, Z_{q a}=1$ denotes that there is an interaction between the user-question or user-answer pair, or the answer belongs to a certain question. 
Specifically, given the user interaction history $\mathcal{H}=\left\{\left(u, \langle q,a\rangle \right) \mid u \in \mathcal{U}, q \in \mathcal{Q}, a \in \mathcal{A}\right\}$, we disentangle it into $\mathcal{H}_q=\left\{\left(u, q\right) \mid u \in \mathcal{U}, q \in \mathcal{Q}\right\}$ and $\mathcal{H}_a=\left\{\left(u, a\right) \mid u \in \mathcal{U}, a \in \mathcal{A}\right\}$ to construct $\mathbf{X}$ and $\mathbf{Y}$. The Q\&A recommendation aims to predict interactions between users and question-answer pairs that are unseen in $\mathcal{H}$.

\section{QAGCF: The Proposed Approach}


In this section, we present the proposed \textbf{Q}uestion \& \textbf{A}nswer \textbf{G}raph \textbf{C}ollaborative \textbf{F}iltering (QAGCF) in three parts. 
We first introduced the method of Question-Answer information disentanglement in Section~\ref{gc}. 
Next, we introduced the polynomial-based graph filters in Section~\ref{pbcf}.
Then we introduced the prediction module based on learned embedding representations in Section~\ref{pred}. Finally, in Section~\ref{ca}, we discuss the complexity analysis of QAGCF.



\subsection{Q-A Information Disentanglement}\label{gc}

As illustrated in Figure~\ref{main_model2}, we initially construct two views: collaborative view and semantic view. The collaboration view is composed of U-Q and U-A graphs to solve the challenge of collaborative information entanglement \textit{\textbf{C1}}, while the semantic view is constructed from Q-A and Q-Q graphs to address the second challenge semantic information entanglement \textit{\textbf{C2} }. 
Then these two views are merged into a global graph.
In subsequent sections, we will delve deeper into the construction of each of these modules.

\subsubsection{Collaborative Information Disentanglement}


Different from previous approaches that only construct bipartite user-item graphs to handle collaborative information, in the context of Q\&A recommendation, it is necessary to construct a collaborative view considering the collaborative relationships between users, questions, and answers. 
We consider disentangling the question-answer pairs into questions and answers, and construct collaborative information between each of them and users separately. Specifically, we construct two bipartite collaboration graphs (U-Q Graph, U-A Graph) to explore the collaborative relationships between users and questions, as well as between users and answers, respectively. 
The adjacency matrices corresponding to the U-Q graph and the U-A graph are as follows:
\begin{equation}
\mathbf{A}^{uq}=\left(\begin{array}{cc}
\mathbf{0} & \mathbf{X} \\
\mathbf{X}^\top & \mathbf{0} \\
\end{array}\right), \quad\quad
\mathbf{A}^{ua}=\left(\begin{array}{cc}
\mathbf{0} & \mathbf{Y} \\
\mathbf{Y}^\top & \mathbf{0} \\
\end{array}\right),
\end{equation}
where $\mathbf{A}^{uq} \in \mathbb{R}^{(M+N) \times (M+N)}$, $\mathbf{A}^{ua} \in \mathbb{R}^{(M+O) \times (M+O)}$.


\subsubsection{Semantic Information Disentanglement}
We construct a semantic view to establish semantic connections both within question-answer pairs and between different pairs. Firstly, we consider the affiliation relationship between each answer and question, which can be represented as a Q-A graph. Here, we establish a correlation matrix to model the semantic correlation between questions and answers. Since each answer is affiliated with only one question, this leads to a dispersed Q-A graph, with each question and its answers forming a star-shaped subgraph with no connections between different questions. To address this, we establish connections between different questions through similarity, creating a Q-Q graph. Combining the Q-A graph with the Q-Q graph allows us to construct relationships between different questions and answers. Next, we will delve into the construction of these two graphs.
\paragraph{Q-A Graph Construction}
We establish the Q-A graph based on the affiliation relationship between each answer and question. Unlike collaborative information, semantic information needs to model the correlation between answers and questions. Specifically, the information transfer between question-answer pairs with lower correlation needs to be weakened. Therefore, we construct a correlation matrix $\mathbf{C}\in \mathbb{R}^{N\times O}$, where each element represents the normalized cosine similarity between questions and answers, in which each entry is calculated as follows:
\begin{equation}
C_{i,j}=\begin{cases}(\mathrm{cos}(\mathbf{e}_{q_i}, \mathbf{e}_{a_j})+1)/2, & Z_{i,j}=1, \\ 0, & \text { otherwise },\end{cases}
\end{equation}
where $\mathbf{e}_{q_i} \in \mathbb{R}^d$ and $\mathbf{e}_{a_j} \in \mathbb{R}^d$ represent the embeddings of question $q_i \in \mathcal{Q}$ and answer $a_j \in \mathcal{A}$, respectively. $d$ is the embedding dimension.
Since the embeddings of questions and answers will be continually updated throughout training, we reconstruct the correlation matrix at the beginning of every epoch.

\paragraph{Q-Q Graph Construction}
To establish connections between different questions, we consider the semantic relationships between them. We compute the pairwise similarity between questions and connect questions with semantic edges when their similarity is relatively high. Specifically, the question-question similarity matrix $\mathbf{S}\in \mathbb{R}^{N\times N}$ is built based on cosine similarity, in which each entry is calculated as follows:
\begin{equation}
S_{i,j}=\mathrm{cos}(\mathbf{e}_{q_i}, \mathbf{e}_{q_j}),    
\end{equation}
where $\mathbf{e}_{q_i} \in \mathbb{R}^d$ and $\mathbf{e}_{q_j} \in \mathbb{R}^d$ represent the embeddings of questions $q_i$ and $q_j$, respectively.

Next, we retain edges with higher similarities on the fully connected Q-Q graph while removing edges between questions with lower similarities to obtain the adjacency matrix $\mathbf{A}^{qq}$ of the Q-Q graph:
\begin{equation}
\label{top-n}
    A^{qq}_{i,j}= \begin{cases}1, & i\neq j\ \ and \  S_{i,j} \in \text { top-}n\left(\mathbf{S}_i\right), \\ 0, & \text { otherwise },\end{cases}
\end{equation}
where the function top-$n$($\cdot$) returns the top-$n$ values of the $i$-th row $\mathbf{S}_i$ of the similarity matrix $\mathbf{S}$. $n$ is a hyperparameter, setting to $\mu N$,where $\mu$ represents the top-$n$ ratio, controling the sparsity of the Q-Q graph. 
Since the question embeddings are continuously updated during training, we reconstruct the Q-Q graph at the start of each epoch.

\subsubsection{Global Graph Construction}
We merge the collaboration and semantic views into a global graph to integrate collaborative and semantic information. The adjacency matrix of the global graph is defined as follows:
\begin{equation}
\label{defa}
\mathbf{A}=\left(\begin{array}{ccc}
\mathbf{0} & \mathbf{X} &  \mathbf{Y}\\
\mathbf{X}^\top & \mathbf{A}^{qq} &  \mathbf{C}\\
\mathbf{Y}^\top & \mathbf{C}^\top & \mathbf{0}\\
\end{array}\right),
\end{equation}
where $\mathbf{A} \in \mathbb{R}^{(M+N+O) \times (M+N+O)}$.
Note that we directly use the correlation matrix $\mathbf{C}$ of the Q-A graph instead of the affiliation matrix $\mathbf{Z}$ to embed the correlation information, which makes subsequent processing more convenient. To ensure training stability, we adopt its normalized form: 
\begin{equation}
    \mathbf{\hat{A}} = \mathbf{D}^{-\frac{1}{2}}\mathbf{A}\mathbf{D}^{-\frac{1}{2}}
\end{equation}
where $\mathbf{D} \in \mathbb{R}^{(M+N+O)\times(M+N+O)}$ is a diagonal matrix, each entry $\mathbf{D}_{ii}$ denotes the number of nonzero entries in the $i$-th row of $\mathbf{A}$.

\begin{figure}[t]
\centering
\includegraphics[width=0.98\columnwidth]{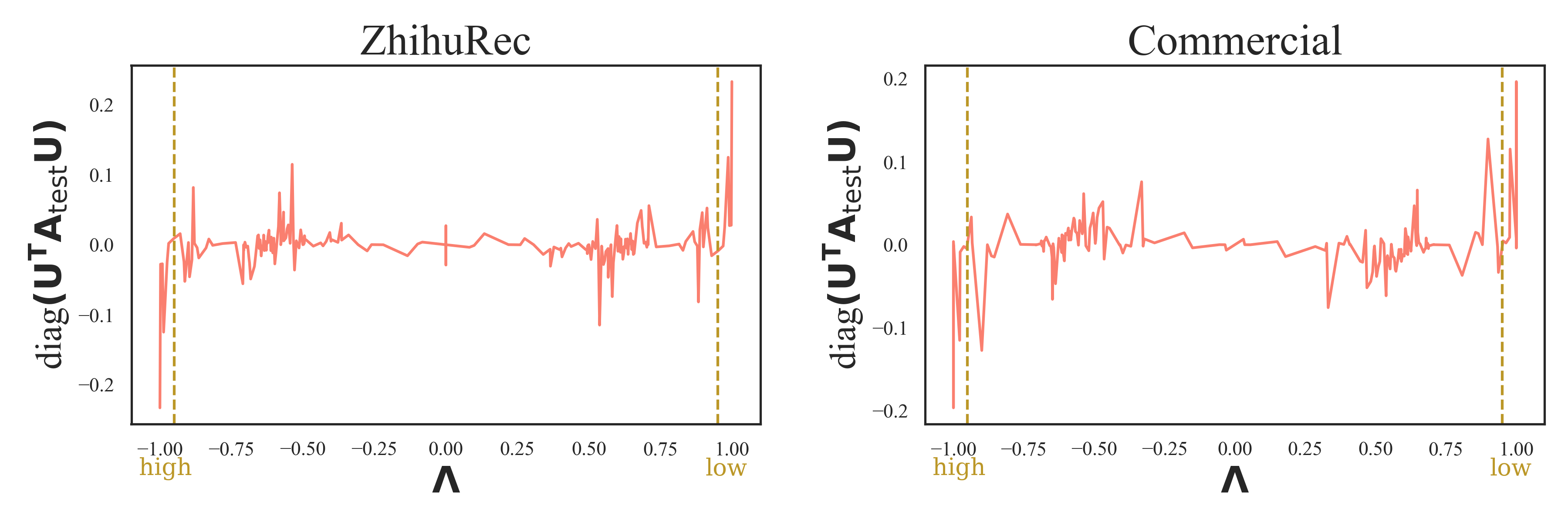}
\caption{Correlation between eigenvalues $\mathbf{\Lambda}$ of $\hat{\mathbf{A}}_{\mathrm{train}}$ and diagonal of $\mathbf{U^\top \mathbf{A}_{\mathrm{test}}U}$ on two Q\&A recommendation datasets. }
\label{signal_plot}
\end{figure}

\subsection{Polynomial-based Graph Filters for Global Graph}\label{pbcf}
In this section, we use polynomial-based graph filters to address the heterophily problem in the global graph.
\subsubsection{Embedding Initialization}
Firstly, we initialize the embeddings as $\mathbf{E}^{(0)} = [\mathbf{E}^{(0)}_{U};\mathbf{E}^{(0)}_{Q};\mathbf{E}^{(0)}_{A}] \in \mathbb{R}^{(M+N+O) \times d}$, where 
$\mathbf{E}^{(0)}_{U}=[\mathbf{e}_{u_1};\mathbf{e}_{u_2};\cdots;\mathbf{e}_{u_M}] \in \mathbb{R}^{M\times d}$, 
$\mathbf{E}^{(0)}_{Q}=[\mathbf{e}_{q_1};\mathbf{e}_{q_2};\cdots;\mathbf{e}_{q_N}] \in \mathbb{R}^{N\times d}$, 
$\mathbf{E}^{(0)}_{A}=[\mathbf{e}_{a_1};\mathbf{e}_{a_2};\cdots;\mathbf{e}_{a_O}] \in \mathbb{R}^{O\times d}$ are the 0-th layer's embeddings of users, questions and answers, respectively.

\subsubsection{Heterophily of the Global Graph}
The global graph contains different types of nodes and edges, resulting in strong heterophily~\cite{yang2020heterogeneous}. Existing graph neural network methods like LightGCN~\cite{he2020lightgcn} mainly focus on low-pass information, making them more suitable for homophilic scenarios~\cite{guo2023manipulating}. 
However, recent studies highlight the importance of high-frequency information in effectively dealing with the complexity of heterophilic graphs~\cite{he2022convolutional, lei2022evennet}. 


We conduct an analysis on ZhihuRec and Commercial datasets from a spectral view to validate the existence of the heterophily problem, as shown in Figure~\ref{signal_plot}.
We can observe a robust linear correlation between the low and high-frequency signals of the normalized adjacency matrix $\hat{\mathbf{A}}_{\mathrm{train}}$ and the adjacency matrix $\mathbf{A}_{\mathrm{test}}$, which are constructed from the interacting data in the training and test sets, respectively. 
However, the mid-frequency signals do not exhibit such a strong linear correlation. Please refer to Appendix~\ref{spv} for further analysis and experimental details.
This observation motivates us to decompose the high and low-frequency signals from the mid-frequency signal and model them separately.


\subsubsection{Graph Signal Frequency Decomposition}
To decompose the mid-frequency signals from the high and low-frequency signals and address the heterophily problem in the global graph, we use the Jacobi polynomial basis
$\mathrm{P}^{\alpha,\beta}_{k}(x)$~\cite{askey1985some}, where $k$ is the order of the polynomial and $\alpha, \beta$ are the parameters that shape the polynomial and weight function. 
The Jacobi polynomial basis has been proven to function like a band-pass filter which can capture the low and high-frequency signals while suppressing the middle-frequency signals~\cite{guo2023manipulating}. 
Specifically, the $k$-th layer's band-stop embeddings which capture the high and low-frequency signals are calculated based on the $k$-th order of the Jacobi polynomial basis:
\begin{equation}
\label{eq:band_stop}
    \mathbf{E}^{(k)}_{\mathrm{band-stop}} = \mathrm{P}^{\alpha,\beta}_{k}(\mathbf{\hat{A}}) \mathbf{E}^{(0)},
\end{equation}
where $\mathbf{E}^{(k)}_{\mathrm{band-stop}} \in \mathbb{R}^{(M+N+O)\times d}$.
Please refer to Appendix~\ref{appendix:Jacobi_Polynomial} for the calculation of the Jacobi polynomial basis $\mathrm{P}^{\alpha,\beta}_{k}(\cdot)$.
For the band-pass embeddings that capture the mid-frequency signals, we subtract it from the original signal for the $k$-th layer as follows:
\begin{equation}
\label{eq:band_pass}
\mathbf{E}^{(k)}_{\mathrm{band-pass}} =\gamma \mathbf{E}^{(0)}-\mathbf{E}^{(k)}_{\mathrm{band-stop}},
\end{equation}
where $\gamma$ is a coefficient that controls the influence of the mid-frequency signals. 
Based on Eq.~\eqref{eq:band_stop} and~\eqref{eq:band_pass}, we can obtain the embedding 
of the $k$-th layer 
\begin{equation}
\label{eq:emb_k}
    \mathbf{E}^{(k)}=[\mathbf{E}^{k}_{\mathrm{band-stop}};\mathbf{E}^{k}_{\mathrm{band-pass}}],
\end{equation}
where $\mathbf{E}^{(k)} \in \mathbb{R}^{(M+N+O)\times 2d}$.
Finally, after passing through $K$ layers, the final embeddings for recommendation are obtained by:
\begin{equation}
\label{eq:final_node}
\mathbf{E} =\frac{1}{K} \sum_{k=1}^K \mathbf{E}^{(k)}.
\end{equation}

\subsection{Prediction and Optimization}
\label{pred}
\subsubsection{Contrastive Learning}
We introduce a contrastive learning loss for learning more robust embeddings during training. Specifically, we implement a noise-based embedding augmentation to the band-stop embedding $\mathbf{e}^{(k)}_{i, \mathrm{band-stop}}$ for node $i$ at layer $k$ ($k>0$):
\begin{equation}
\mathbf{e}_{i,\mathrm{band-stop}}^{(k)\prime}=\mathbf{e}^{(k)}_{i,\mathrm{band-stop}}+\Delta,
\end{equation}
where the added noise vectors $\Delta \in \mathbb{R}^d$ is required that:
\begin{equation}
\Delta=\omega \odot \operatorname{sign}\left(\mathbf{e}^{(k)}_{i,\text{band-stop}}\right), \omega \in \mathbb{R}^d \sim U(0,1),    
\end{equation}
which forces the noise not result in a large deviation. 
Then according to Eq.~\eqref{eq:band_pass}, the augmented band-pass embedding of node $i$ at layer $k$ is:
\begin{equation}
\mathbf{e}_{i,\mathrm{band-pass}}^{(k)\prime}=\gamma\mathbf{e}_i^{(0)}-\mathbf{e}^{(k)\prime}_{i,\mathrm{band-stop}}-\Delta.
\end{equation}
Then, based on Eq.~\eqref{eq:emb_k} and~\eqref{eq:final_node}, we can obtain the augmented embedding $\mathbf{e}^{(k)\prime}_{i} = [\mathbf{e}_{i,\mathrm{band-stop}}^{(k)\prime};\mathbf{e}_{i,\mathrm{band-pass}}^{(k)\prime}]$ of node $i$ at the $k$-th layer as well as the final embedding $e^{\prime}_{i}$.
Next, we define the cross-layer contrastive loss as follows:
\begin{equation}
\left\{\begin{array}{l}
\mathcal{L}^{C}_u = -\sum_{i \in \mathcal{U}}\log \frac{\exp \left(\mathbf{e}_i^{\prime \top} \mathbf{e}_i^{(l)\prime} / \tau\right)}{\sum_{j \in \mathcal{U}_{\mathrm{neg}}} \exp \left(\mathbf{e}_i^{\prime \top} \mathbf{e}_j^{(l)\prime} / \tau\right)}, \\
\mathcal{L}^{C}_q = -\sum_{i \in \mathcal{Q}}\log \frac{\exp \left(\mathbf{e}_i^{\prime \top} \mathbf{e}_i^{(l)\prime} / \tau\right)}{\sum_{j \in \mathcal{Q}_{neg}} \exp \left(\mathbf{e}_i^{\prime \top} \mathbf{e}_j^{(l)\prime} / \tau\right)}, \\
\mathcal{L}^{C}_a = -\sum_{i \in \mathcal{A}}\log \frac{\exp \left(\mathbf{e}_i^{\prime \top} \mathbf{e}_i^{(l)\prime} / \tau\right)}{\sum_{j \in \mathcal{A}_{\mathrm{neg}}} \exp \left(\mathbf{e}_i^{\prime \top} \mathbf{e}_j^{(l)\prime} / \tau\right)},
\end{array}\right.
\end{equation}
where $\tau$ is the temperature. $l$ denotes the layer used for contrast with the final embedding, which is set to $1$ in our model. $\mathcal{U}_{neg}$, $\mathcal{A}_{neg}$, $\mathcal{Q}_{neg}$ are the randomly sampled in-batch negatives. 
$\mathcal{L}^{C}_u$, $\mathcal{L}^{C}_q$ and $\mathcal{L}^{C}_a$ are the contrastive losses for users, questions and answers, respectively.
The total contrastive loss is computed as:
\begin{equation}
\mathcal{L}^{C} = \mathcal{L}^{C}_u + \mathcal{L}^{C}_q + \mathcal{L}^{C}_a.
\end{equation}
The objective of this contrastive loss is to bring representations of the same node closer across different layers, while pushing representations of different nodes across different layers farther apart, thus alleviating the potential oversmoothing issue~\cite{rusch2023survey} caused by excessive layers and learning more robust representations.

\subsubsection{Training}
Firstly, we compute the inner product of user-question and user-answer according to the augmented embedding $\mathbf{e}_{i}^{\prime}$, then combine them through a weighted summation to get the final prediction:
\begin{equation}
\hat{y}_{u, q, a}={\mathrm{e}_u^{\prime\top}} \mathrm{e}_a^{\prime}+\lambda_1{\mathrm{e}_u^{\prime\top}} \mathrm{e}_q^{\prime}.
\label{predeq}
\end{equation}
where $\lambda_1$ is the coefficient that controls the influence of question, The widely used Bayesian Personalized Ranking (BPR) loss~\cite{rendle2012bpr} is then adopted:
\begin{equation}
\mathcal{L}^{B P R}=\sum_{\left(u, \langle q,a\rangle, \langle q^{\prime}, a^{\prime}\rangle\right) \in O}-\ln \sigma\left(\hat{y}_{u, q, a}-\hat{y}_{u, q^{\prime}, a^{\prime}}\right) ,  
\end{equation}
where $\sigma(\cdot)$ is the sigmoid function, $O=\big\{\left(u, \langle q,a\rangle, \langle q^{\prime}, a^{\prime}\rangle\right) \mid X_{uq}=1, Y_{ua} = 1, Z_{qa}=1, X_{uq^{\prime}}=0, Y_{ua^{\prime}}=0, Z_{q^{\prime}a^{\prime}}=1\big\}$ denotes the training data, and $\langle q^{\prime}, a^{\prime}\rangle$ denotes the sampled pair that user $u$ has not interacted with. 

The final loss $\mathcal{L}$ include the BPR loss $\mathcal{L}^{B P R}$ and the contrastive loss $\mathcal{L}^C$:
\begin{equation}
\mathcal{L}=\mathcal{L}^{B P R}+\lambda_2\mathcal{L}^C+\lambda_3\|\Theta\|_2^2,    
\end{equation}
where $\lambda_2$ and $\lambda_3$ are the hyperparameters that control the influence of the contrastive loss and $L_2$ regularization, respectively. And $\Theta$ are the parameters of our model.

\subsection{Complexity Analysis}\label{ca}
Regarding the space complexity, the parameters of QAGCF only include three sets of embeddings: $\mathbf{E}_U^{(0)}$, $\mathbf{E}_Q^{(0)}$, $\mathbf{E}_A^{(0)}$. The total space complexity of QAGCF is $\mathcal{O}((M+N+O)d)$. 

As for the time complexity, the computational cost of QAGCF primarily comes from Q-A correlation matrix computation, Q-Q graph construction, tripartite graph learning, and contrastive learning. The time complexity of Q-A correlation matrix and Q-Q graph construction in QAGCF is $O((N+O+\mu N\log(N))Nds)$, where $\mu$ is the defined $\text{top-}n$ ratio. The time complexity for graph learning is $\mathcal{O}((|E_{UQ}| + |E_{UA}| + |E_{QA}| + \mu N^2)Kds\frac{|E_{UA}|}{B})$, where $|E_{UQ}|$, $|E_{UA}|$, $|E_{QA}|$ are the number of edges in the U-Q graph, U-A graph, Q-A graph, respectively, $K$ is the number of propagation layers, $s$ is the number of epoch, $B$ is the batch size. The time complexity for computing the contrastive loss is $\mathcal{O}(B^2 d)$.
\begin{table}[t]
\centering
\caption{Dataset Statistics.}
\begin{tabular}{l|rrrr}
\hline Dataset & \#Users & \#Questions & \#Answers & \# Interactions \\
\hline
\hline
ZhihuRec & 7,975 & 12,612 & 31,025 & 265,971 \\
Commercial & 40,204 & 31,710 & 101,056 & 953,212 \\
\hline
\end{tabular}
\label{tab:stat}
\end{table}
\begin{table*}
\centering
  \caption{Performance Comparison of Different Recommendation Models. The best result is bolded and the runner-up is underlined. * means improvements over the second-best methods are significant (\textit{t}-test, \textit{p}-value < 0.05).}
  \label{tab:main}
\begin{tabular}{cc|ccccc|ccccc}
\hline
\multirow{2}{*}{Category} & \multirow{2}{*}{Model} & \multicolumn{5}{c}{ZhihuRec} & \multicolumn{5}{c}{Commercial} \\
\cline{3-7} \cline{8-12}
& & Recall & MRR & NDCG & Hit & Precision & Recall & MRR & NDCG & Hit & Precision \\
\hline
\hline
\multirow{2}{*}{Non-GNN} & BPRMF & $0.0363$ & $0.0324$ & $0.0236$ & $0.0934$ & $0.0098$ & $0.0496$ & $0.0281$ & $0.0289$ & $0.0728$ & $0.0074$ \\
& NeuMF& $0.0285$ & $0.0257$ & $0.0185$ & $0.0760$ & $0.0079$ & $0.0256$ & $0.0142$ & $0.0146$ & $0.0387$ & $0.0039$ \\
\hline
\multirow{4}{*}{GNN} & NGCF & $0.0460$ & $0.0395$ & $0.0292$ & $0.1163$ & $0.0124$ & $0.0532$ & $0.0297$ & $0.0307$ & $0.0773$ & $0.0079$ \\
& LightGCL & $0.0440$ & $0.0409$ & $0.0292$ & $0.1131$ & $0.0112$ & $0.0396$ & $0.0227$ & $0.0234$ & $0.0575$ & $0.0580$ \\
& HMLET & $0.0511$ & $0.0466$ & $0.0335$ & $0.1290$ & $0.0140$ & $0.0618$ & $0.0350$ & $0.0361$ & $0.0896$ & $0.0092$ \\
& LightGCN & $0.0481$ & $0.0441$ & $0.0317$ & $0.1206$ & $0.0130$ & $0.0604$ & $0.0343$ & $0.0353$ & $0.0880$ & $0.0090$ \\
\hline
\multirow{3}{*}{GNN+PbGF} & JGCF & $0.0510$ & $0.0469$ & $0.0339$ & $0.1295$ & $0.0140$ & $0.0626$ & $0.0361$ & $0.0367$ & $0.0914$ & $0.0094$ \\
& JGCF\_L & $0.0502$ & $0.0466$ & $0.0334$ & $0.1264$ & $0.0137$ & $0.0621$ & $0.0354$ & $0.0364$ & $0.0906$ & $0.0093$ \\
& JGCF\_C & $0.0512$ & $0.0473$ & $0.0338$ & $0.1293$ & $0.0141$ & $0.0643$ & $0.0366$ & $0.0375$ & $0.0935$ & $0.0097$ \\

\hline
\multirow{3}{*}{GNN+CL} & SGL & $0.0495$ & $0.0465$ & $0.0328$ & $0.1279$ & $0.0140$ & $0.0639$ & $0.0370$ & $0.0377$ & $0.0932$ & $0.0096$ \\
& SimGCL & $0.0532$ & $0.0471$ & $0.0342$ & $0.1299$ & $0.0140$ & $0.0700$ & $0.0405$ & $0.0413$ & $0.1018$ & $0.0105$ \\
& XSimGCL & $\underline{0.0562}$ & $\underline{0.0503}$ & $\underline{0.0366}$ & $\underline{0.1403}$ & $\underline{0.0155}$ & $\underline{0.0756}$ & $\underline{0.0447}$ & $\underline{0.0453}$ & $\underline{0.1093}$ & $\underline{0.0113}$ \\
\hline
OURS& QAGCF & $\mathbf{0.0592}^*$ & $\mathbf{0.0543}^*$ & $\mathbf{0.0391}^*$ & $\mathbf{0.1484}^*$ & $\mathbf{0.0160}^*$ & $\textbf{0.0795}^*$ & $\textbf{0.0461}^*$ & $\textbf{0.0472}^*$ & $\textbf{0.1144}^*$ & $\textbf{0.0119}^*$ \\
\hline \hline
\multicolumn{2}{c|}{\%Improv.}  & $\textbf{+5.34\%}$ & $\textbf{+7.95\%}$ & $\textbf{+6.83\%}$ & $\textbf{+5.77\%}$ & $\textbf{+3.23\%}$ & $\textbf{+5.16\%}$ & $\textbf{+3.13\%}$ & $\textbf{+4.19\%}$ & $\textbf{+4.67\%}$ & $\textbf{+5.31\%}$ \\
\hline
\end{tabular}
\end{table*}

\section{Experiments}
To verify the effectiveness of QAGCF, we conduct extensive experiments and report detailed analysis results.
\subsection{Experimental Settings}
\subsubsection{Datasets} To evaluate the performance of the proposed QAGCF, we use two Q\&A community datasets to conduct experiments: ZhihuRec and a Commercial dataset. The statistics of the datasets are summarized in Table~\ref{tab:stat}. 
\begin{itemize} [leftmargin=*]
    \item \textbf{ZhihuRec}\footnote{\url{https://github.com/THUIR/ZhihuRec-Dataset}}~\cite{hao2021large} is collected from a large-scale knowledge-acquisition platform (Zhihu), including original data with question information, answer information and user profiles. We selected the 1M version of the dataset, filtered out answers not affiliated with any questions, and retained only positive interactions.
    \item \textbf{Commercial} is constructed based on the user behavior logs on a public Q\&A platform, recording the interactions of users regarding questions and answers. This dataset includes over $40,000$ users and $1$ million interactions.
\end{itemize}

Following existing works~\cite{ye2023towards}, we divide each dataset into three parts, \emph{i.e.}, training/validation/test sets according to a ratio of 8:1:1. We employ a uniform sampling approach, selecting one negative item for each positive instance to form the training set.
\subsubsection{Baselines} We compare QAGCF with four categories of baseline models: Non-GNN, GNN, GNN+Polynomial-based Graph Filter and GNN+Contrastive Learning.

\textbf{Non-GNN}:
\textbf{BPRMF~\cite{rendle2012bpr}} is a matrix factorization algorithm that optimizes Bayesian Personalized Ranking.
\textbf{NeuMF~\cite{he2017neural}} combines matrix factorization and neural networks to model linear and non-linear relationships, improving recommendation accuracy.

\textbf{GNN}:
\textbf{NGCF~\cite{wang2019neural}} utilizes a user-item bipartite graph and employs Graph Neural Networks (GNNs) to incorporate high-order relations.
\textbf{LightGCL~\cite{cai2023lightgcl}} utilizes singular value decomposition for contrastive augmentation, enabling unconstrained semantic refinement and global collaboration in graph learning.
\textbf{HMLET~\cite{kong2022linear}} is a hybrid collaborative filtering model that combines linear and nonlinear propagation and dynamically selects between these modes.
\textbf{LightGCN~\cite{he2020lightgcn}} simplifies the design of GCN to make it more concise and appropriate for recommendation.

\textbf{GNN+Polynomial-based Graph Filter}:
\textbf{JGCF~\cite{guo2023manipulating}} applies Jacobi polynomial basis to capture spectral features. We also compared its variants using two other polynomial basis (Legendre, Chebyshev) named \textbf{JGCF\_L} and \textbf{JGCF\_C} respectively. 

\textbf{GNN+Contrastive Learning}:
\textbf{SGL~\cite{wu2021self}} applies self-supervised contrastive learning to the LightGCN model for graph augmentation. We adopt SGL-ED as the instantiation of SGL.
\textbf{SimGCL~\cite{yu2022graph}} proposes an effective recommendation model without graph augmentation, enhancing data through the addition of uniform random noises to user and course representations.
\textbf{XSimGCL~\cite{yu2023xsimgcl}} is an improvement over SimGCL, achieves similar results by combining task recommendation and comparison, reducing model complexity through noise enhancement and interlayer contrastive learning.

\begin{table*}[htbp]
\centering
  \caption{Ablation Study and Analysis of Graph Construction in QAGCF.}
  \label{tab:ablation}
\begin{tabular}{l|ccccc|ccccc}
\hline
\multirow{2}{*}{Model} & \multicolumn{5}{c}{ZhihuRec} & \multicolumn{5}{c}{Commercial} \\
\cline{2-6} \cline{7-11}
& Recall & MRR & NDCG & Hit & Precision & Recall & MRR & NDCG & Hit & Precision \\
\hline
\hline
QAGCF\_UA & $0.0565$ & $0.0524$ & $0.0374$ & $0.1433$ & $0.0154$ & $0.0776$ & $0.0450$ & $0.0461$ & $0.1118$ & $0.0116$\\
QAGCF\_UA+UQ & $0.0579$ & $0.0540$ & $0.0387$ & $0.1472$ & $0.0158$ & $0.0788$ & $0.0461$ & $0.0470$ & $0.1128$ & $0.0117$\\
\hdashline
QAGCF w/o PbGF & $0.0575$ & $0.0510$ & $0.0372$ & $0.1441$ & $0.0158$ & $0.0786$ & $0.0456$ & $0.0466$ & $0.1132$ & $0.0117$ \\
QAGCF w/o CL & $0.0527$ & $0.0474$ & $0.0349$ & $0.1329$ & $0.0143$ & $0.0710$ & $0.0408$ & $0.0419$ & $0.1023$ & $0.0106$ \\
\hdashline
QAGCF\_U(A+Q) & $0.0566$ & $0.0535$ & $0.0383$ & $0.1418$ & $0.0154$ & $0.0778$ & $0.0455$ & $0.0465$ & $0.1122$ & $0.0117$\\
\hdashline
QAGCF & $0.0592$ & $0.0543$ & $0.0391$ & $0.1484$ & $0.0160$ & $0.0795$ & $0.0461$ & $0.0472$ & $0.1144$ & $0.0119$\\
\hline
\end{tabular}
\end{table*}

\subsubsection{Evaluation Metrics} To evaluate the performance of top-$K$ recommendation, we adopt the widely used metrics: Recall@$K$, MRR@$K$, NDCG@$K$, Hit@$K$ and Precision@$K$. In this context, we set $K = 10$ and present the average scores on the testing set. Following ~\cite{lin2022improving, he2020lightgcn, wu2021self}, we adopt the full-ranking strategy~\cite{zhao2020revisiting}, which ranks all the candidate items that the user has not interacted with.
\subsubsection{Implementation Details} 
Our implementation of QAGCF and all baseline models is conducted using RecBole-GNN\footnote{\url{https://github.com/RUCAIBox/RecBole-GNN}}~\cite{zhao2021recbole, zhao2022recbole2}, a unified open-source framework for developing and reproducing recommendation algorithms. To ensure a fair comparison, we optimize all methods with the Adam~\cite{kingma2014adam} optimizer and carefully search hyperparameters for all baselines. The batch size is fixed at $2048$, and parameters are initialized using the Xavier~\cite{glorot2010understanding} distribution, with an embedding size set to $64$. Validation set testing is performed every $10$ epochs, incorporating early stopping with patience of $10$ epochs to prevent overfitting, where MRR@$10$ serves as the indicator. We fine-tune hyperparameters for optimal performance.

\subsection{Overall Performance}

Table~\ref{tab:main} presents the performance comparison of the proposed QAGCF and other baseline methods on two datasets. From the table, we observe several key insights: 

(1)~Compared to traditional methods like BPRMF, methods based on Graph Neural Networks (GNNs) perform better as they encode higher-order information of the graph into their representations. 

(2)~Compared to graph collaborative filtering baseline models like LightGCN, JGCF based on polynomial-based graph filters demonstrates superior performance, highlighting the importance of capturing mid- and high-frequency information in the graph. 

(3)~Among all graph collaborative filtering baselines, XSimGCL exhibits the best performance on most datasets, emphasizing the effectiveness of contrastive learning in enhancing recommendation performance. However, XSimGCL does not construct collaborative and semantic views regarding questions and answers for users, this makes its performance inferior to QAGCF.

(4)~Ultimately, the proposed QAGCF consistently outperforms baseline models. In contrast to these baselines, our work delves into the intricate relationships among users, questions, and answers, including both collaborative and semantic connections. By employing polynomial-based graph filters alongside contrastive learning loss, we achieve a more effective learning of node representations, setting our approach apart in its ability to 
understand and utilize the specificities of Q\&A recommendation.

\begin{figure}[t]
\centering
\includegraphics[width=0.56\columnwidth]{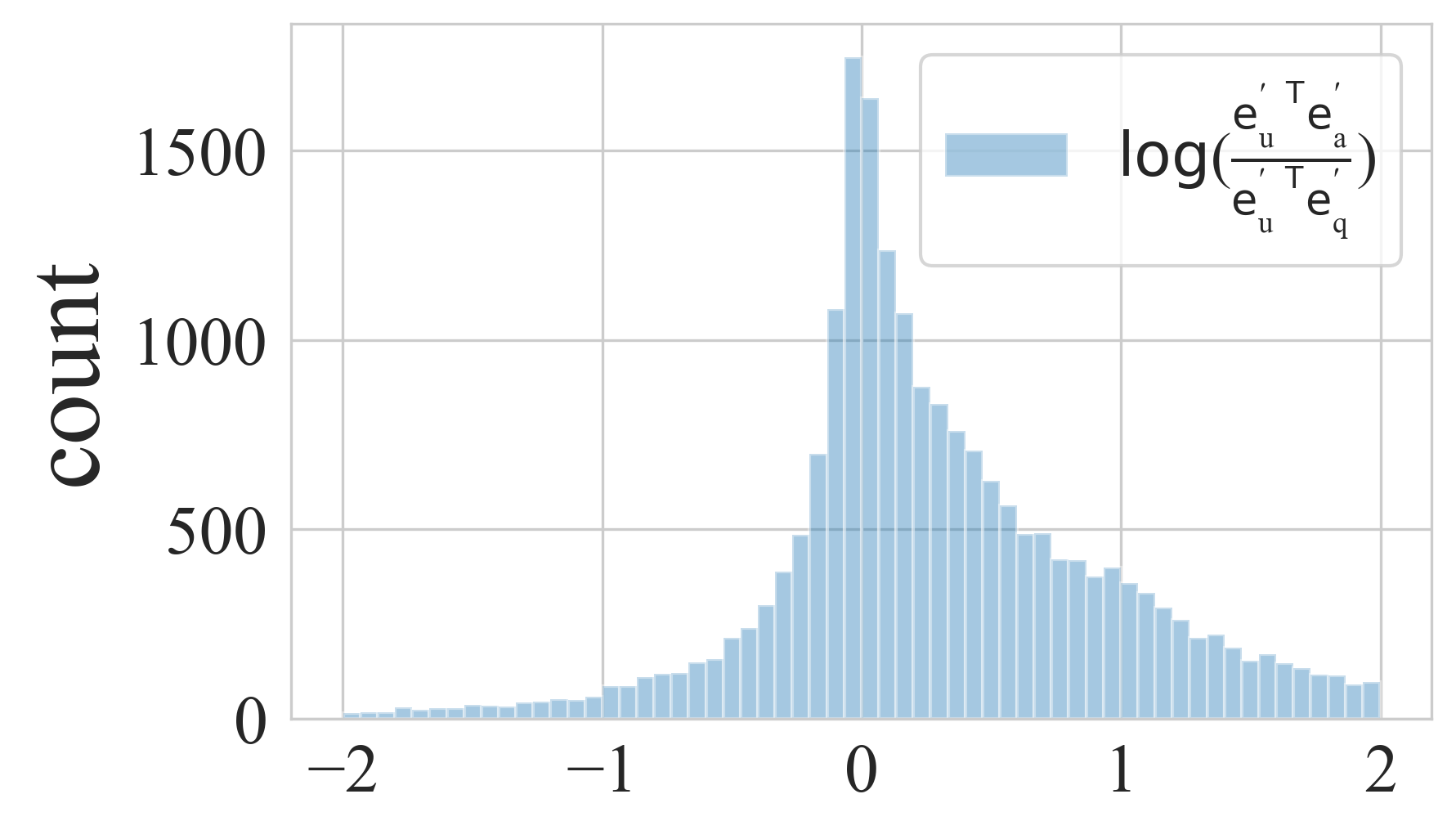}
\caption{Validation of QAGCF's Disentanglement of Collaborative Information. Logarithmic operation is applied for better visualization.}
\label{dis}
\vspace{-0.5cm}
\end{figure}




\subsection{Ablation Study}

To further evaluate the key components of QAGCF, we conducted a series of ablation studies as shown in Table~\ref{tab:ablation}.

\subsubsection{Effectiveness of Graph Construction.}
To evaluate the contribution of the graph construction in QAGCF, we separately retained the U-A graph, U-A and U-Q graph, named QAGCF\_UA and QAGCF\_UA+UQ, respectively. It can be observed in Table~\ref{tab:ablation} that the recommendation performance of these two models progressively increases, yet remains below that of the complete QAGCF with all graphs retained. This indicates the importance and rationality of constructing both the collaboration and semantic views.

\subsubsection{Effectiveness of Polynomial-based Graph Filter}
To evaluate the performance of polynomial-based graph filters, we first removed the polynomial-based graph filters and directly trained using a LightGCN-based approach, named QAGCF w/o PbGF. It can be observed in Table~\ref{tab:ablation} that removing the polynomial-based graph filters resulted in a significant performance drop for QAGCF. 
This demonstrates that decomposing low and high frequency from mid-frequency information using polynomial-based graph filters is helpful for better handling graphs with strong heterophily.

\subsubsection{Effectiveness of Contrastive Learning}
To evaluate whether the contrastive learning loss contributes to the performance, we removed the contrastive learning module, named QAGCF w/o CL. It can be observed in Table~\ref{tab:ablation} that QAGCF experienced a substantial performance decline, demonstrating the necessity of using contrastive learning to learn a more robust embedding.


\begin{figure}[t]
\centering
\includegraphics[width=0.33\textwidth]{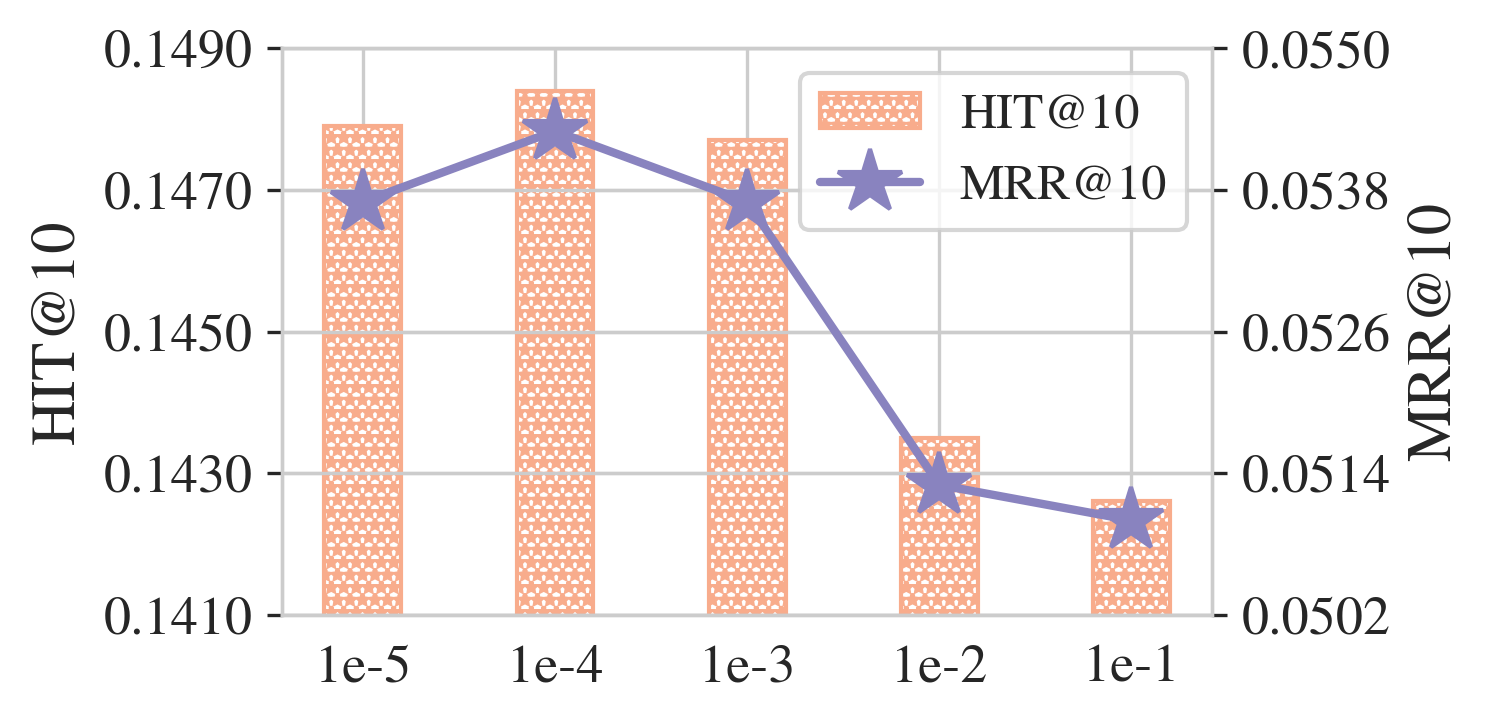}
\caption{Sensitivity Analysis of $\mu$ in Q-Q Graph Construction.}
\label{ana1}
\vspace{-0.5cm}
\end{figure}

\subsection{Detailed empirical analysis}

We conducted more detailed experiments on ZhihuRec to show how and why QAGCF can improve the recommendation performance.


\begin{figure*}[t]
\centering
\subfigure[Performance comparison w.r.t. different $\alpha$] {\includegraphics[width=.65\columnwidth]{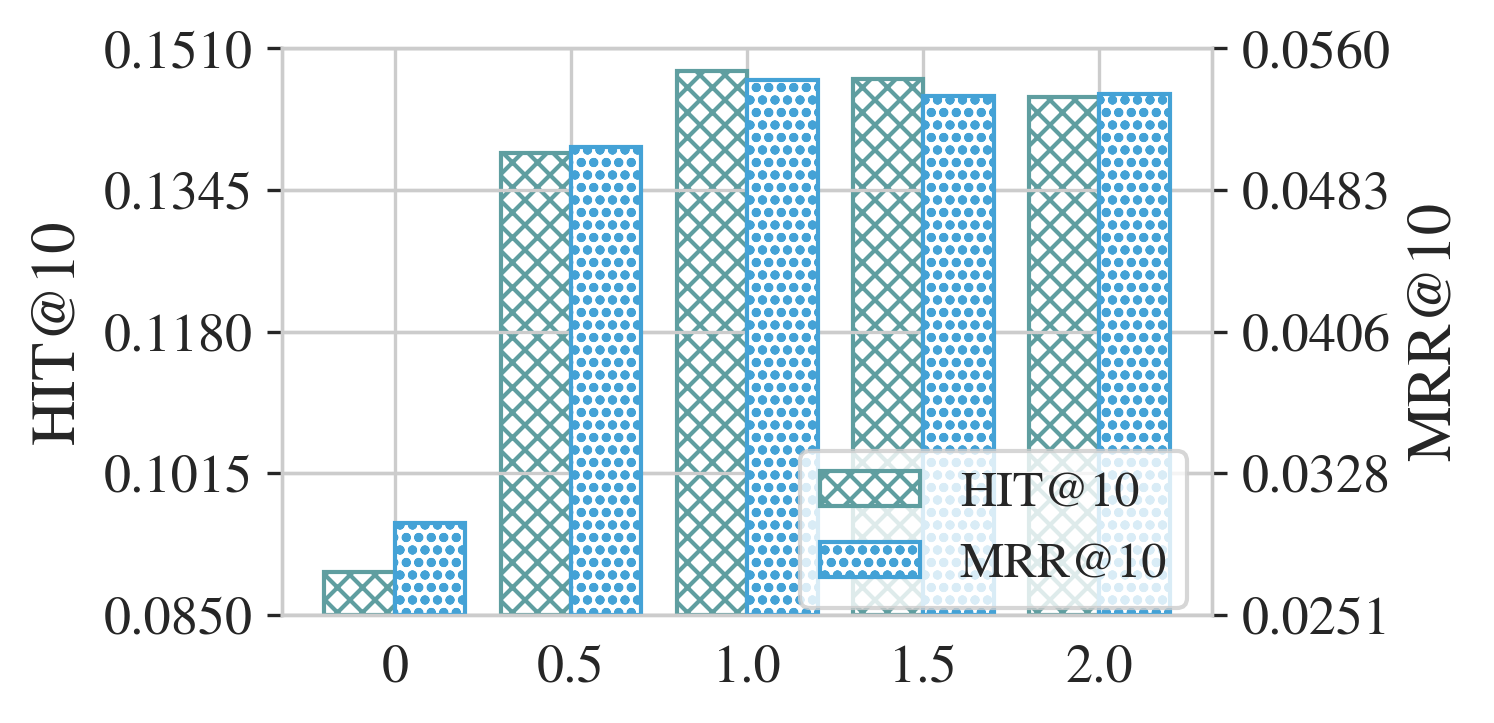}}
\subfigure[Performance comparison w.r.t. different $\beta$] {\includegraphics[width=.65\columnwidth]{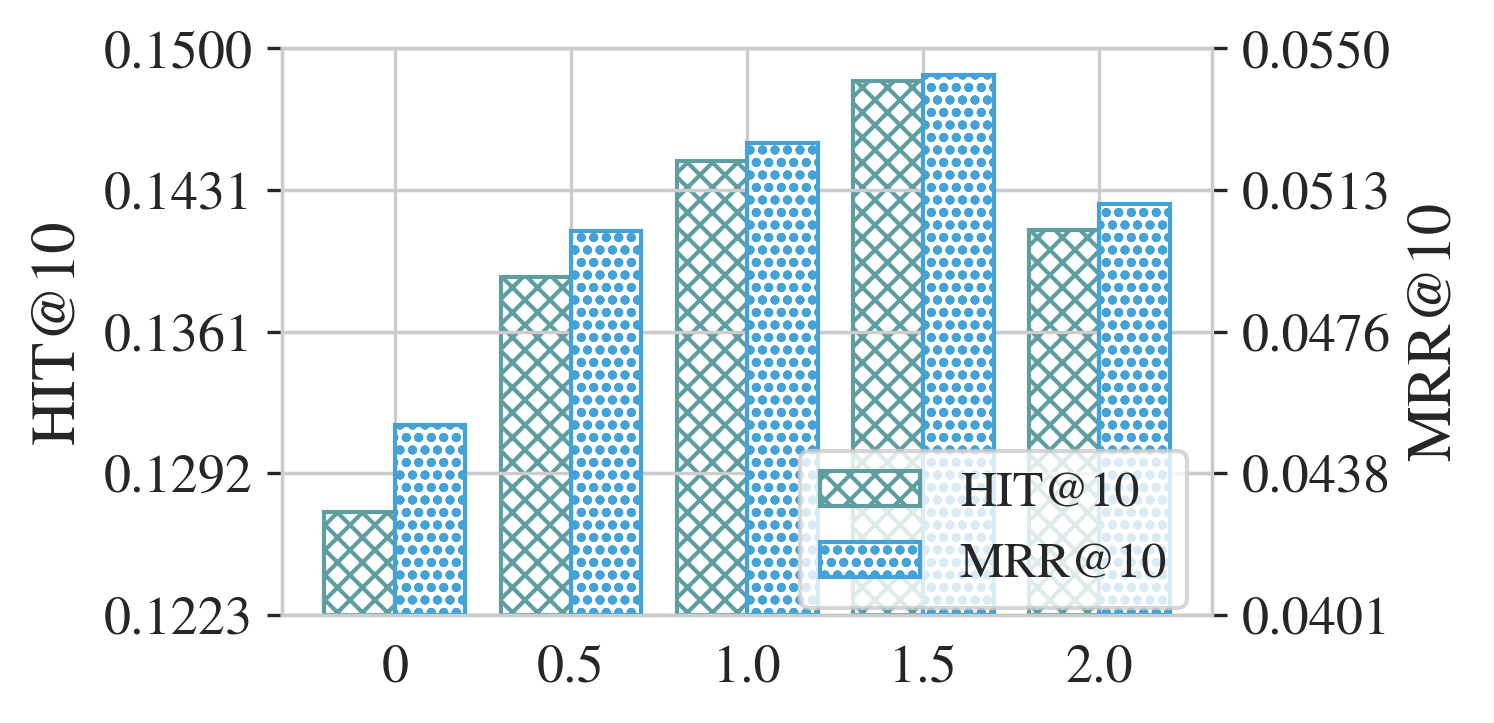}}
\subfigure[Performance comparison w.r.t. different $\gamma$] {\includegraphics[width=.65\columnwidth]{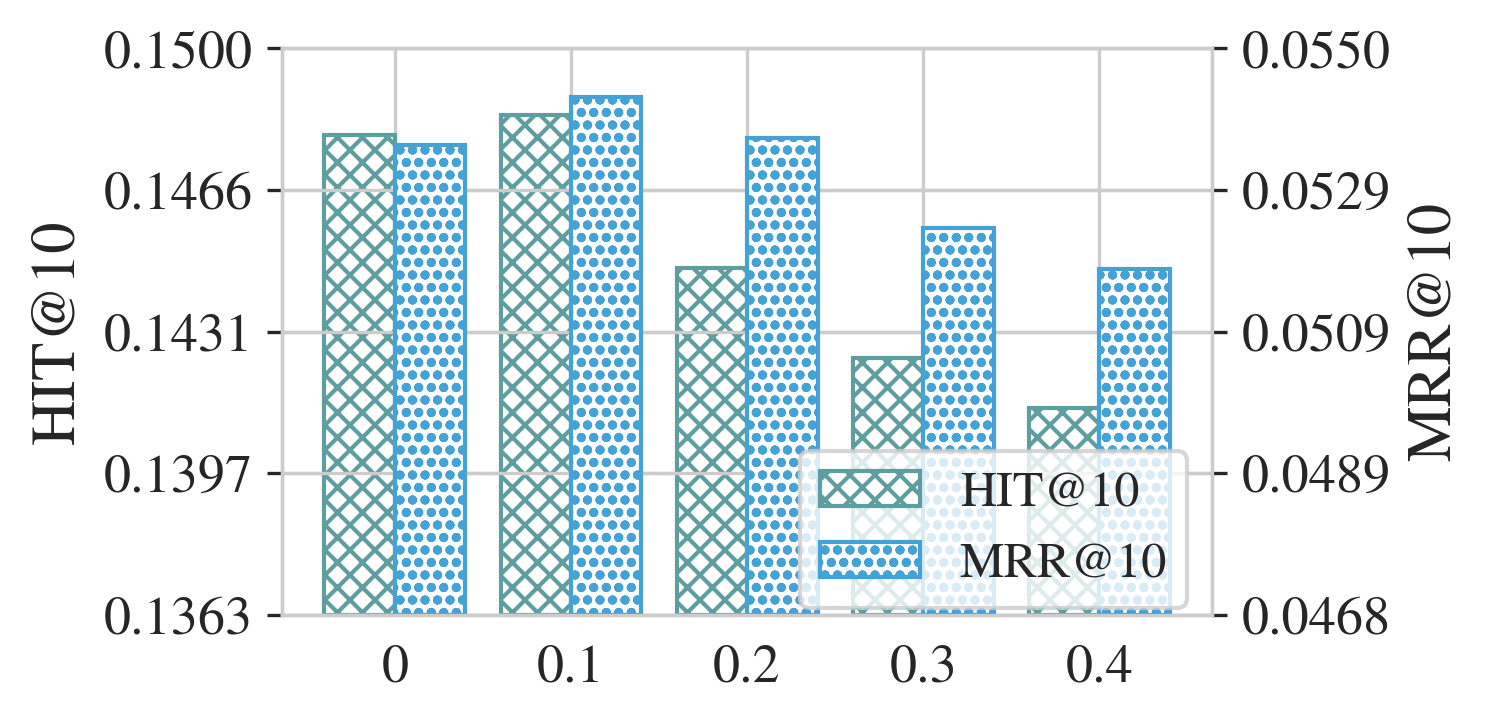}}
\vspace{-5px}
\caption{Sensitivity Analysis of $\alpha$, $\beta$ and $\gamma$ in Polynomial-based Graph Filters.}
\label{ana2}
\vspace{-0.5cm}
\end{figure*}

\begin{figure*}[t]
\centering
\subfigure[Performance comparison w.r.t. different $\lambda_2$] {\includegraphics[width=.65\columnwidth]{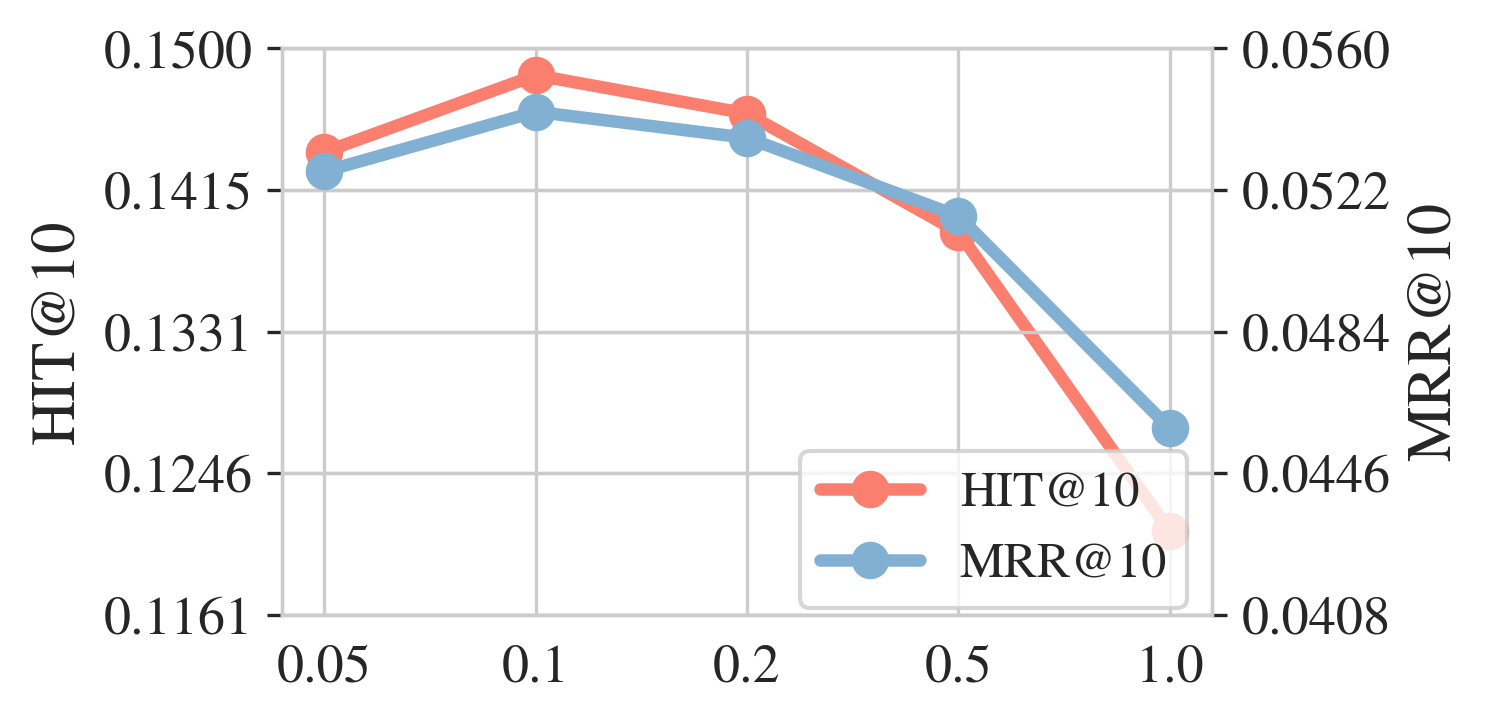}}
\subfigure[Performance comparison w.r.t. different $\omega$] {\includegraphics[width=.65\columnwidth]{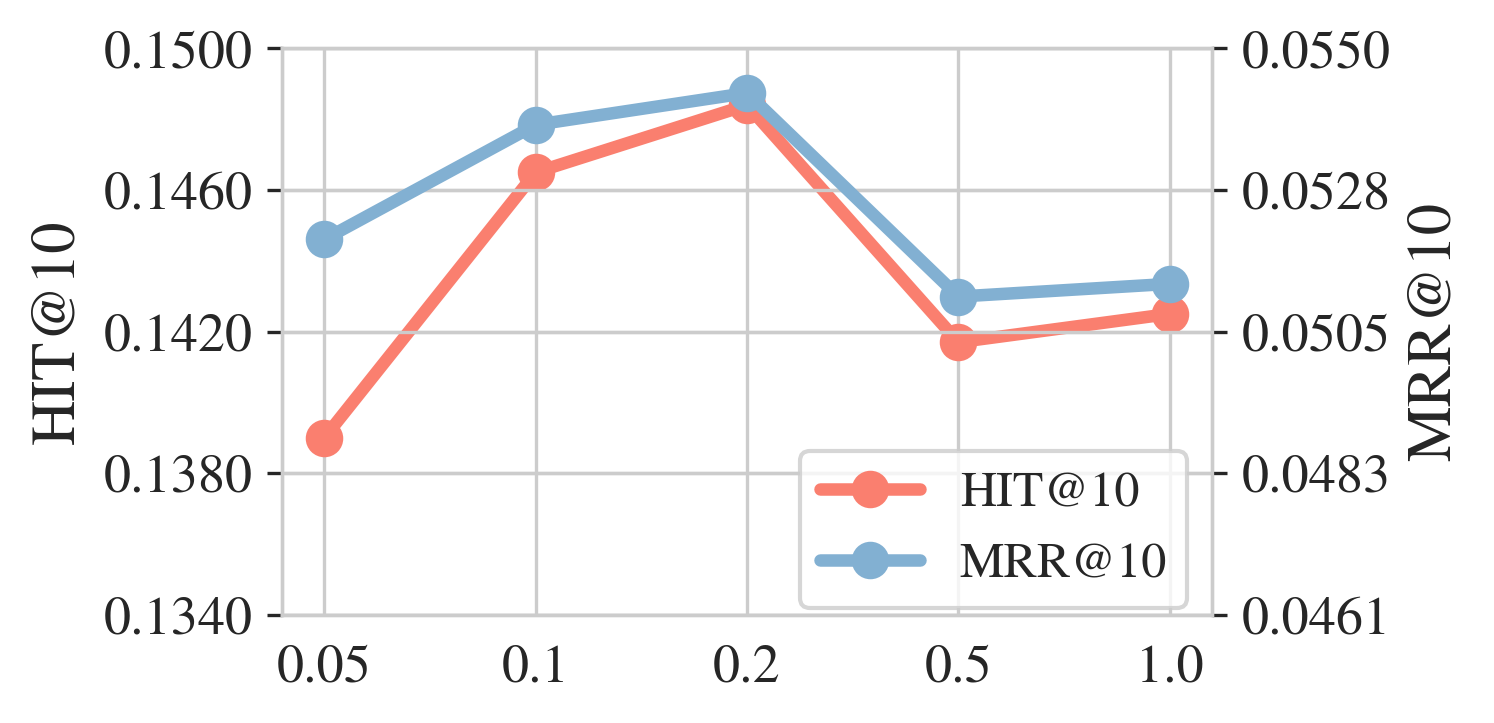}}
\subfigure[Performance comparison w.r.t. different $\tau$] {\includegraphics[width=.65\columnwidth]{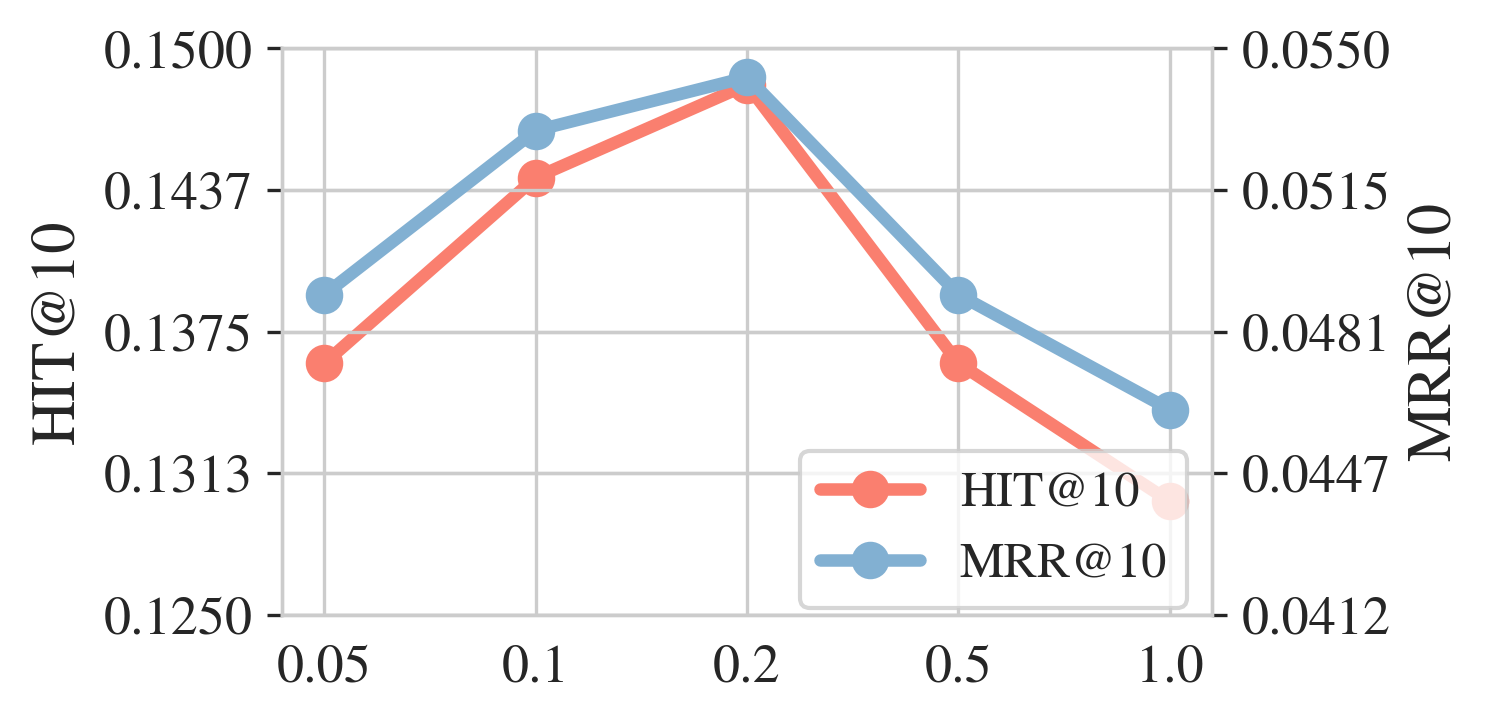}}
\vspace{-5px}
\caption{Sensitivity Analysis of $\lambda_2$, $\omega$ and $\tau$ in Contrastive Learning Module.}
\label{ana3}
\vspace{-0.5cm}
\end{figure*}

\subsubsection{Discussion of Collaborative Information Disentanglement}
This section investigates the effectiveness of QAGCF in collaborative information disentanglement. We address the challenge of collaborative information entanglement by constructing two bipartite graphs, U-Q and U-A, respectively, within the collaborative view. In Eq.~\eqref{predeq} we perform dot product and linear weighting between the user embedding $e_u$, question embedding $e_q$, and answer embedding $e_a$. Subsequently, we compute the ratio of the dot product of each user-question interaction and the dot product of each user-answer interaction on the test set, and plot it in Figure~\ref{dis}. It is observed that this ratio can be very low, indicating that the user is significantly more interested in the answer than the question; conversely, this ratio can also be very high, suggesting that the user is significantly more interested in the question than the answer. 
This experiment demonstrates that by separately constructing the U-Q and U-A graphs, we can capture the different impacts that questions and answers have on user clicks. It shows that our model successfully disentangles the collaborative information between users and questions as well as between users and answers.

\subsubsection{Analysis of Graph Construction in QAGCF}
The core component of QAGCF revolves around the construction of two collaborative graphs: the U-Q graph and the U-A graph, along with two semantic graphs: the Q-A graph and the Q-Q graph. 
To illustrate the importance of such graph construction, we compared QAGCF with the method that directly considers questions and answers as a single item, which is denoted as QAGCF\_U(A+Q).
For QAGCF\_U(A+Q), we directly add the embedding representations of questions to the embedding representations of answers, resulting in the initial embedding representation of answers in the bipartite graph. The design of other modules remains the same as in QAGCF.
The comparison with QAGCF\_UA and QAGCF\_U(A+Q) in Table~\ref{tab:ablation} reveals that adding the embeddings of questions to answers improves the model's performance on both datasets. However, there is a significant performance gap compared to QAGCF, which constructs all four graphs. This highlights the importance of constructing both collaborative and semantic views.

\subsubsection{Impact of $\mu$}
We analyzed the influence of the parameter $\mu$ on the construction of the Q-Q graph in Figure~\ref{ana1}. $\mu$ affects the sparsity of the Q-Q graph, where the sparsity of the Q-Q graph is calculated as $1-\mu$. The larger the value of $\mu$, the denser the Q-Q graph. It can be observed in Figure~\ref{ana1} that the optimal choice for $\mu$ is $1e\text{-}4$, resulting in a sparsity of the Q-Q graph of $99.99\%$. When $\mu=1e\text{-}5$, only the most similar question is selected for each question on ZhihuRec, resulting in an overly sparse graph that leads to poor recommendation performance. As $\mu$ increases from $1e\text{-}3$ to $1e\text{-1}$, the performance decreases as the Q-Q graph becomes denser.

\subsubsection{Impact of $\alpha$, $\beta$ and $\gamma$}
We analysis the impact of parameters $\alpha$, $\beta$, and $\gamma$ in the polynomial-based graph filters in Figure~\ref{ana2}, where $\alpha$ and $\beta$ respectively control the influence of low-frequency and high-frequency signals in the band-stop filter, and $\gamma$ controls the magnitude of the mid-frequency signals in the band-pass filter. We chose $\alpha$ and $\beta$ both from $[0, 0.5, 1.0, 1.5, 2.0]$, and $\gamma$ from $[0, 0.1, 0.2, 0.3, 0.4]$ for experimentation. It can be observed from Figure~\ref{ana2} that $\alpha=1.0$, $\beta=1.5$, $\gamma=0.1$ are the optimal choices. We observe that the parameter $\beta$ plays a more crucial role in influencing the performance of QAGCF, as low-frequency signals contain abundant information about similar users and items~\cite{guo2023manipulating}. However, we should not set $\alpha$ too large, as emphasizing low-frequency signals excessively may lead to over-smoothing~\cite{guo2023manipulating}. Regarding parameter $\alpha$, performance remains high within the range of $1.0\sim 2.0$. This is because high-frequency signals enable the model to learn differences between similar nodes, thereby enriching its knowledge. Additionally, introducing high-frequency signals may introduce some randomness, reducing the likelihood of overfitting. Clearly, emphasizing both low and high-frequency signals is beneficial for learning in the highly heterogeneous graph of QAGCF, while appropriately suppressing mid-frequency signals also contributes to better recommendation performance.

\subsubsection{Impact of $\lambda_2$, $\omega$ and $\tau$}
We analysis the impact of parameters $\lambda_2$, $\omega$, and $\tau$ in the contrastive learning module in Figure~\ref{ana3}. For fair comparison, when studying $\lambda_2$, we fix $\omega=0.2$ and $\tau=0.2$; when studying $\omega$, we fix $\lambda_2=0.1$ and $\tau=0.2$; when studying $\tau$, we fix $\lambda_2=0.1$ and $\omega=0.2$. It can be observed from the figure that the performance of the model is more sensitive to variations in $\lambda_2$ and $\tau$. The optimal parameter settings on ZhihuRec are observed to be $\lambda_2=0.1$, $\omega=0.2$, and $\tau=0.2$. The results demonstrate that QAGCF is quite sensitive to the hyperparameters and both too large and too small values of these parameters lead to a decrease in recommendation performance.

\subsubsection{Impact of $\lambda_1$}
\begin{figure}[t]
\centering
\includegraphics[width=0.66\columnwidth]{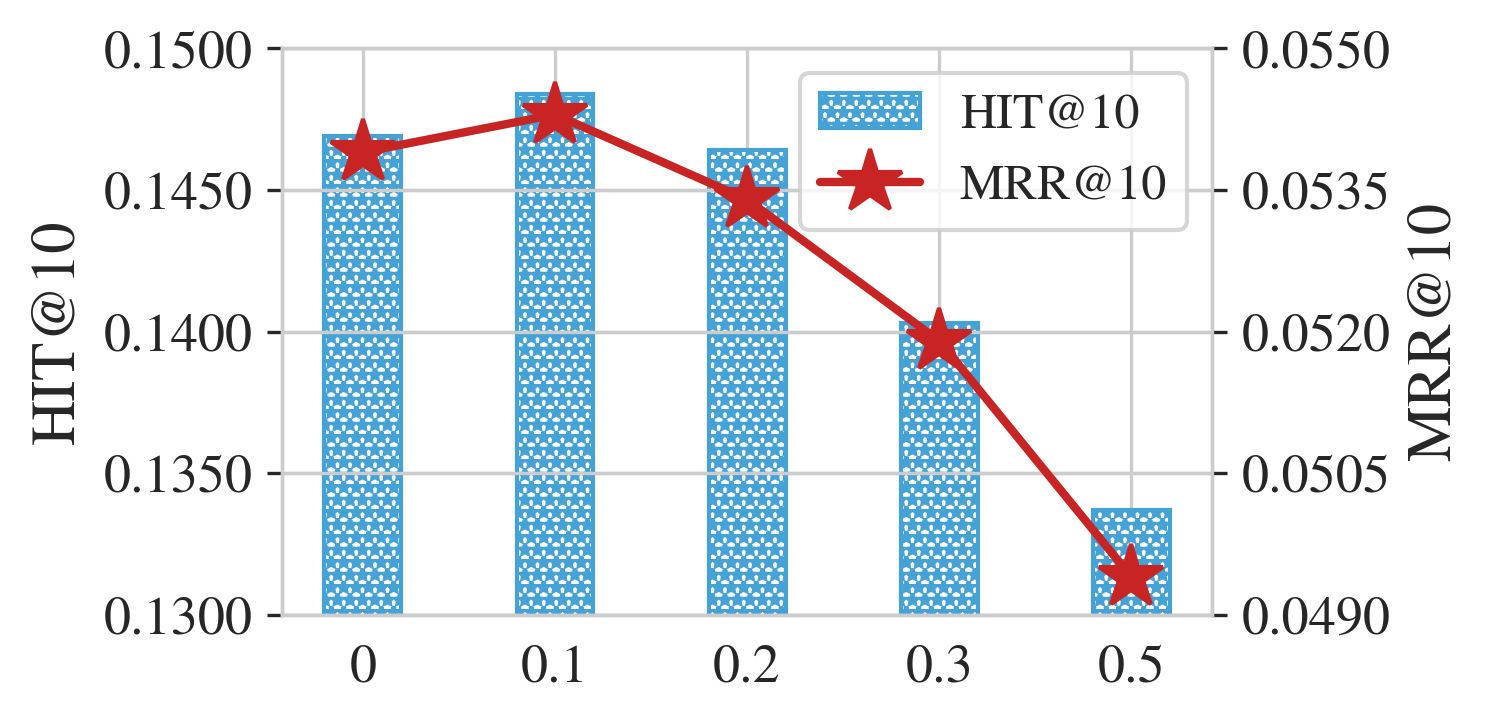}
\caption{Sensitivity Analysis of $\lambda_1$ in Prediction Module.}
\label{ana4}
\vspace{-0.5cm}
\end{figure}

We analysis the impact of parameters $\lambda_1$ in the prediction module in Figure~\ref{ana4}. $\lambda_1$ is the weighted coefficient of $\mathrm{e}_u^{\mathrm{~T}} \mathrm{e}_q$ in $\hat{y}_{u, q, a}$. A smaller value of $\lambda_1$ indicates a smaller influence of $\mathrm{e}_q$ on the predicted score. From the graph, we can observe that the optimal choice for $\lambda_1$ is 0.1. Therefore, $\hat{y}_{u, q, a}$ is more influenced by $\mathrm{e}_u^{\mathrm{~T}} \mathrm{e}_a$. This is because there is a hierarchical relationship between questions and answers, and the answers can more accurately reflect the user's interest compared to questions.

\section{Related Work}

\subsection{Research on Q\&A platforms}
Existing research on Q\&A platforms has primarily focused on finding potential responders to questions or providing high-quality answers to users' queries. Efforts include developing algorithms to calculate user authority in specific fields, using social network analysis, and the HITS and PageRank algorithms to identify experts~\cite{jurczyk2007discovering, shen2009recommending, zhang2007expertise, zhao2014expert, zhao2016expert}. Other studies have concentrated on reducing response wait times by finding similar questions or relevant answers within large archives by employing various information retrieval techniques~\cite{jeon2006framework, hammond1997question, zhang2020answer, kundu2020preference, kundu2019formulation, matsubara2020reranking}. Research has explored community dynamics like motivations for contribution and how user reputation impacts perceived answer quality~\cite{liu2013integrating,tausczik2011predicting}. 
Some have also examined question recommendation through classification or ranking models, considering user interests and feedback~\cite{dror2011want,guo2008tapping, qu2009probabilistic}.

\subsection{Graph-based collaborative filtering.}

Graph-based collaborative filtering, empowered by Graph Neural Networks (GNNs) like NGCF, LightGCN, and LCF \cite{wang2019neural,he2020lightgcn,yu2022graph}, has revolutionized recommendation systems. These models excel in aggregating user-item graph information, enhancing recommendation performance. LightGCN's streamlined architecture has inspired efficient recommendation approaches. Moreover, CL-based models like SGL and MHCN \cite{wu2021self,yu2021self} have emerged, along with self-supervised learning techniques in SimGCL and XSimGCL \cite{yu2022graph,yu2023xsimgcl}, further improving recommendation generalization. These advancements signify a transition from traditional methods to graph-based models, better capturing intricate user-item interactions.

\section{Conclusion}
In this work, we introduced a novel graph neural network model, QAGCF, for recommendation task on Q\&A platforms, utilizing a multi-view approach to disentangle collaborative information and semantic information. Our findings indicate that QAGCF surpasses existing models in accuracy, providing a promising direction for enhancing user experience on Q\&A platforms. Despite its effectiveness, the exploration of graph neural networks in Q\&A recommendation remains nascent, with potential for future advancements in modeling techniques and application scopes.
\appendix

\section{Appendix}
\subsection{Analysis on Q\&A Recommdendation}
\label{spv}
This section analyses the Q\&A recommendation graph collaborative filtering from a spectral transformation view~\cite{guo2023manipulating}, which transforms the recommendation system's challenge into predicting unobserved interactions by learning a function $f(\cdot)$ that maps the normalized adjacency matrix $\hat{\mathbf{A}}_{\mathrm{train}}$ constructed from the interacting data in training set to the adjacency matrix $\mathbf{A}_{\mathrm{test}}$ constructed from the test set. 
Through eigen-decomposition $\hat{\mathbf{A}}_{\mathrm{train}}=\mathbf{U}\boldsymbol{\Lambda}\mathbf{U}^T$ and optimization focused on minimizing the Frobenius norm, this approach simplifies to a one-dimensional least squares problem: ${\operatorname{minimize}} \sum_i\left(f(\boldsymbol{\Lambda}_{ii})-\mathbf{U}_{\cdot i}^\top \mathbf{A}_{\mathrm{test}} \mathbf{U}_{\cdot i}\right)^2$, aiming to accurately predict future user-item interactions.

The same description is used in our Q\&A recommendation task, where the adjacency matrix $\mathbf{A}_{\mathrm{train}}$ and $\mathbf{A}_{\mathrm{test}}$ are both calculated by Eq.~\eqref{defa}. 
We conducted experiments on ZhihuRec-1M and our Commercial dataset. First, we sampled a small portion of highly correlated interaction data from the dataset for easier eigen-decomposition. Subsequently, we divided the dataset into a training set and a test set in an $8:2$ ratio to construct $\mathbf{A}_{\mathrm{train}}$ and $\mathbf{A}_{\mathrm{test}}$. After experimental analysis, we observed the relationship between the diagonal elements of $\mathbf{\Lambda}$ and $\mathbf{U^\top A_{\mathrm{test}}U}$ as depicted in Figure~\ref{signal_plot}. The experimental results are similar to the phenomena described in~\cite{guo2023manipulating}: the low-frequency component (values around 1.0) and the high-frequency component (values around -1.0) are positively correlated with the diagonal of $\mathbf{U^\top A_{\mathrm{test}}U}$, while the mid-frequency component (values around 0.0) is randomly scattered and has a lower correlation with the signal on the test graph.
This observation motivates us to decompose the high and low-frequency signals from the mid-frequency signal and model them separately.

\subsection{Jacobi Polynomial Basis}
\label{appendix:Jacobi_Polynomial}
Jacobi polynomial basis ~\cite{askey1985some} can be used as band-stop filters in signal processing, suppressing mid-frequency signals while enhancing low-frequency and high-frequency signals. The formula for Jacobi polynomial basis is given by:
\begin{equation}
    \mathrm{P}_k^{\alpha, \beta}(x) = \frac{(-1)^k}{2^k k!} (1 - x)^{-\alpha} (1 + x)^{-\beta} \frac{d^k}{dx^k} \left[ (1 - x)^{k+\alpha} (1 + x)^{k+\beta} \right].
    \label{eq:jacobi}
\end{equation}
where $k$ is the order of the polynomial, $\alpha$ and $\beta$ are parameters that shape the polynomial and weight function. 
These polynomials are orthogonal to the weight function $(1 - x)^{\alpha} (1 + x)^{\beta}$ on the interval $[-1, 1]$.
When \(k = 0\) or \(k = 1\), the Jacobi polynomial simplifies to:
\begin{equation}
    \begin{aligned}
        & \mathrm{P}_0^{\alpha, \beta}(x) = 1,\\
& \mathrm{P}_1^{\alpha, \beta}(x) = \frac{ (\alpha + \beta + 2)x + (\alpha - \beta) }{2}.
    \end{aligned}
\end{equation}
When \(k \geq 2\), the Jacobi polynomials can be computed using the recurrence relation:
\begin{equation}
\mathrm{P}_k^{\alpha, \beta}(x) = \left(\theta_k x + \theta_k^{\prime}\right) \mathrm{P}_{k-1}^{\alpha, \beta}(x) - \theta_k^{\prime \prime} \mathrm{P}_{k-2}^{\alpha, \beta}(x),
\end{equation}
where
\begin{equation}
\begin{aligned}
& \theta_k = \frac{(2 k + \alpha + \beta)(2 k + \alpha + \beta - 1)}{2 k(k + \alpha + \beta)}, \\
& \theta_k^{\prime} = \frac{(2 k + \alpha + \beta - 1)(\alpha^2 - \beta^2)}{2 k(k + \alpha + \beta)(2 k + \alpha + \beta - 2)}, \\
& \theta_k^{\prime \prime} = \frac{(k + \alpha - 1)(k + \beta - 1)(2 k + \alpha + \beta)}{k(k + \alpha + \beta)(2 k + \alpha + \beta - 2)} .
\end{aligned}
\end{equation}



\bibliographystyle{ACM-Reference-Format}
\bibliography{ref}


\begin{thebibliography}{46}


\ifx \showCODEN    \undefined \def \showCODEN     #1{\unskip}     \fi
\ifx \showDOI      \undefined \def \showDOI       #1{#1}\fi
\ifx \showISBNx    \undefined \def \showISBNx     #1{\unskip}     \fi
\ifx \showISBNxiii \undefined \def \showISBNxiii  #1{\unskip}     \fi
\ifx \showISSN     \undefined \def \showISSN      #1{\unskip}     \fi
\ifx \showLCCN     \undefined \def \showLCCN      #1{\unskip}     \fi
\ifx \shownote     \undefined \def \shownote      #1{#1}          \fi
\ifx \showarticletitle \undefined \def \showarticletitle #1{#1}   \fi
\ifx \showURL      \undefined \def \showURL       {\relax}        \fi
\providecommand\bibfield[2]{#2}
\providecommand\bibinfo[2]{#2}
\providecommand\natexlab[1]{#1}
\providecommand\showeprint[2][]{arXiv:#2}

\bibitem[Askey and Wilson(1985)]%
        {askey1985some}
\bibfield{author}{\bibinfo{person}{Richard Askey} {and} \bibinfo{person}{James~Arthur Wilson}.} \bibinfo{year}{1985}\natexlab{}.
\newblock \bibinfo{booktitle}{\emph{Some basic hypergeometric orthogonal polynomials that generalize Jacobi polynomials}}. Vol.~\bibinfo{volume}{319}.
\newblock \bibinfo{publisher}{American Mathematical Soc.}
\newblock


\bibitem[Cai et~al\mbox{.}(2023)]%
        {cai2023lightgcl}
\bibfield{author}{\bibinfo{person}{Xuheng Cai}, \bibinfo{person}{Chao Huang}, \bibinfo{person}{Lianghao Xia}, {and} \bibinfo{person}{Xubin Ren}.} \bibinfo{year}{2023}\natexlab{}.
\newblock \showarticletitle{LightGCL: Simple Yet Effective Graph Contrastive Learning for Recommendation}.
\newblock \bibinfo{journal}{\emph{arXiv preprint arXiv:2302.08191}} (\bibinfo{year}{2023}).
\newblock


\bibitem[Dai et~al\mbox{.}(2024)]%
        {dai2024model}
\bibfield{author}{\bibinfo{person}{Sunhao Dai}, \bibinfo{person}{Ninglu Shao}, \bibinfo{person}{Jieming Zhu}, \bibinfo{person}{Xiao Zhang}, \bibinfo{person}{Zhenhua Dong}, \bibinfo{person}{Jun Xu}, \bibinfo{person}{Quanyu Dai}, {and} \bibinfo{person}{Ji-Rong Wen}.} \bibinfo{year}{2024}\natexlab{}.
\newblock \showarticletitle{Modeling User Attention in Music Recommendation}. In \bibinfo{booktitle}{\emph{2024 IEEE 40th International Conference on Data Engineering (ICDE)}}.
\newblock


\bibitem[Dror et~al\mbox{.}(2011)]%
        {dror2011want}
\bibfield{author}{\bibinfo{person}{Gideon Dror}, \bibinfo{person}{Yehuda Koren}, \bibinfo{person}{Yoelle Maarek}, {and} \bibinfo{person}{Idan Szpektor}.} \bibinfo{year}{2011}\natexlab{}.
\newblock \showarticletitle{I want to answer; who has a question? Yahoo! answers recommender system}. In \bibinfo{booktitle}{\emph{Proceedings of the 17th ACM SIGKDD international conference on Knowledge discovery and data mining}}. \bibinfo{pages}{1109--1117}.
\newblock


\bibitem[Glorot and Bengio(2010)]%
        {glorot2010understanding}
\bibfield{author}{\bibinfo{person}{Xavier Glorot} {and} \bibinfo{person}{Yoshua Bengio}.} \bibinfo{year}{2010}\natexlab{}.
\newblock \showarticletitle{Understanding the difficulty of training deep feedforward neural networks}. In \bibinfo{booktitle}{\emph{Proceedings of the thirteenth international conference on artificial intelligence and statistics}}. JMLR Workshop and Conference Proceedings, \bibinfo{pages}{249--256}.
\newblock


\bibitem[Guo et~al\mbox{.}(2023)]%
        {guo2023manipulating}
\bibfield{author}{\bibinfo{person}{Jiayan Guo}, \bibinfo{person}{Lun Du}, \bibinfo{person}{Xu Chen}, \bibinfo{person}{Xiaojun Ma}, \bibinfo{person}{Qiang Fu}, \bibinfo{person}{Shi Han}, \bibinfo{person}{Dongmei Zhang}, {and} \bibinfo{person}{Yan Zhang}.} \bibinfo{year}{2023}\natexlab{}.
\newblock \showarticletitle{On Manipulating Signals of User-Item Graph: A Jacobi Polynomial-based Graph Collaborative Filtering}.
\newblock \bibinfo{journal}{\emph{arXiv preprint arXiv:2306.03624}} (\bibinfo{year}{2023}).
\newblock


\bibitem[Guo et~al\mbox{.}(2008)]%
        {guo2008tapping}
\bibfield{author}{\bibinfo{person}{Jinwen Guo}, \bibinfo{person}{Shengliang Xu}, \bibinfo{person}{Shenghua Bao}, {and} \bibinfo{person}{Yong Yu}.} \bibinfo{year}{2008}\natexlab{}.
\newblock \showarticletitle{Tapping on the potential of q\&a community by recommending answer providers}. In \bibinfo{booktitle}{\emph{Proceedings of the 17th ACM conference on Information and knowledge management}}. \bibinfo{pages}{921--930}.
\newblock


\bibitem[Hammond et~al\mbox{.}(1997)]%
        {hammond1997question}
\bibfield{author}{\bibinfo{person}{K Hammond}, \bibinfo{person}{R Burke}, \bibinfo{person}{V Kulyukin}, \bibinfo{person}{S Lytinen}, \bibinfo{person}{N Tomuro}, {and} \bibinfo{person}{S Schoenberg}.} \bibinfo{year}{1997}\natexlab{}.
\newblock \showarticletitle{Question Answering from Frequently-Asked Question Files: Experiences with the FAQ Finder System}.
\newblock  (\bibinfo{year}{1997}).
\newblock


\bibitem[Hao et~al\mbox{.}(2021)]%
        {hao2021large}
\bibfield{author}{\bibinfo{person}{Bin Hao}, \bibinfo{person}{Min Zhang}, \bibinfo{person}{Weizhi Ma}, \bibinfo{person}{Shaoyun Shi}, \bibinfo{person}{Xinxing Yu}, \bibinfo{person}{Houzhi Shan}, \bibinfo{person}{Yiqun Liu}, {and} \bibinfo{person}{Shaoping Ma}.} \bibinfo{year}{2021}\natexlab{}.
\newblock \showarticletitle{A large-scale rich context query and recommendation dataset in online knowledge-sharing}.
\newblock \bibinfo{journal}{\emph{arXiv preprint arXiv:2106.06467}} (\bibinfo{year}{2021}).
\newblock


\bibitem[He et~al\mbox{.}(2022)]%
        {he2022convolutional}
\bibfield{author}{\bibinfo{person}{Mingguo He}, \bibinfo{person}{Zhewei Wei}, {and} \bibinfo{person}{Ji-Rong Wen}.} \bibinfo{year}{2022}\natexlab{}.
\newblock \showarticletitle{Convolutional neural networks on graphs with chebyshev approximation, revisited}.
\newblock \bibinfo{journal}{\emph{Advances in Neural Information Processing Systems}}  \bibinfo{volume}{35} (\bibinfo{year}{2022}), \bibinfo{pages}{7264--7276}.
\newblock


\bibitem[He et~al\mbox{.}(2020)]%
        {he2020lightgcn}
\bibfield{author}{\bibinfo{person}{Xiangnan He}, \bibinfo{person}{Kuan Deng}, \bibinfo{person}{Xiang Wang}, \bibinfo{person}{Yan Li}, \bibinfo{person}{Yongdong Zhang}, {and} \bibinfo{person}{Meng Wang}.} \bibinfo{year}{2020}\natexlab{}.
\newblock \showarticletitle{Lightgcn: Simplifying and powering graph convolution network for recommendation}. In \bibinfo{booktitle}{\emph{Proceedings of the 43rd International ACM SIGIR conference on research and development in Information Retrieval}}. \bibinfo{pages}{639--648}.
\newblock


\bibitem[He et~al\mbox{.}(2017)]%
        {he2017neural}
\bibfield{author}{\bibinfo{person}{Xiangnan He}, \bibinfo{person}{Lizi Liao}, \bibinfo{person}{Hanwang Zhang}, \bibinfo{person}{Liqiang Nie}, \bibinfo{person}{Xia Hu}, {and} \bibinfo{person}{Tat-Seng Chua}.} \bibinfo{year}{2017}\natexlab{}.
\newblock \showarticletitle{Neural collaborative filtering}. In \bibinfo{booktitle}{\emph{Proceedings of the 26th international conference on world wide web}}. \bibinfo{pages}{173--182}.
\newblock


\bibitem[Huzhang et~al\mbox{.}(2021)]%
        {huzhang2021aliexpress}
\bibfield{author}{\bibinfo{person}{Guangda Huzhang}, \bibinfo{person}{Zhenjia Pang}, \bibinfo{person}{Yongqing Gao}, \bibinfo{person}{Yawen Liu}, \bibinfo{person}{Weijie Shen}, \bibinfo{person}{Wen-Ji Zhou}, \bibinfo{person}{Qing Da}, \bibinfo{person}{Anxiang Zeng}, \bibinfo{person}{Han Yu}, \bibinfo{person}{Yang Yu}, {et~al\mbox{.}}} \bibinfo{year}{2021}\natexlab{}.
\newblock \showarticletitle{AliExpress Learning-To-Rank: Maximizing online model performance without going online}.
\newblock \bibinfo{journal}{\emph{IEEE Transactions on Knowledge and Data Engineering}} (\bibinfo{year}{2021}).
\newblock


\bibitem[Jeon et~al\mbox{.}(2006)]%
        {jeon2006framework}
\bibfield{author}{\bibinfo{person}{Jiwoon Jeon}, \bibinfo{person}{W~Bruce Croft}, \bibinfo{person}{Joon~Ho Lee}, {and} \bibinfo{person}{Soyeon Park}.} \bibinfo{year}{2006}\natexlab{}.
\newblock \showarticletitle{A framework to predict the quality of answers with non-textual features}. In \bibinfo{booktitle}{\emph{Proceedings of the 29th annual international ACM SIGIR conference on Research and development in information retrieval}}. \bibinfo{pages}{228--235}.
\newblock


\bibitem[Jurczyk and Agichtein(2007)]%
        {jurczyk2007discovering}
\bibfield{author}{\bibinfo{person}{Pawel Jurczyk} {and} \bibinfo{person}{Eugene Agichtein}.} \bibinfo{year}{2007}\natexlab{}.
\newblock \showarticletitle{Discovering authorities in question answer communities by using link analysis}. In \bibinfo{booktitle}{\emph{Proceedings of the sixteenth ACM conference on Conference on information and knowledge management}}. \bibinfo{pages}{919--922}.
\newblock


\bibitem[Kingma and Ba(2014)]%
        {kingma2014adam}
\bibfield{author}{\bibinfo{person}{Diederik~P Kingma} {and} \bibinfo{person}{Jimmy Ba}.} \bibinfo{year}{2014}\natexlab{}.
\newblock \showarticletitle{Adam: A method for stochastic optimization}.
\newblock \bibinfo{journal}{\emph{arXiv preprint arXiv:1412.6980}} (\bibinfo{year}{2014}).
\newblock


\bibitem[Kong et~al\mbox{.}(2022)]%
        {kong2022linear}
\bibfield{author}{\bibinfo{person}{Taeyong Kong}, \bibinfo{person}{Taeri Kim}, \bibinfo{person}{Jinsung Jeon}, \bibinfo{person}{Jeongwhan Choi}, \bibinfo{person}{Yeon-Chang Lee}, \bibinfo{person}{Noseong Park}, {and} \bibinfo{person}{Sang-Wook Kim}.} \bibinfo{year}{2022}\natexlab{}.
\newblock \showarticletitle{Linear, or non-linear, that is the question!}. In \bibinfo{booktitle}{\emph{Proceedings of the fifteenth ACM international conference on web search and data mining}}. \bibinfo{pages}{517--525}.
\newblock


\bibitem[Kundu and Mandal(2019)]%
        {kundu2019formulation}
\bibfield{author}{\bibinfo{person}{Dipankar Kundu} {and} \bibinfo{person}{Deba~Prasad Mandal}.} \bibinfo{year}{2019}\natexlab{}.
\newblock \showarticletitle{Formulation of a hybrid expertise retrieval system in community question answering services}.
\newblock \bibinfo{journal}{\emph{Applied Intelligence}}  \bibinfo{volume}{49} (\bibinfo{year}{2019}), \bibinfo{pages}{463--477}.
\newblock


\bibitem[Kundu et~al\mbox{.}(2020)]%
        {kundu2020preference}
\bibfield{author}{\bibinfo{person}{Dipankar Kundu}, \bibinfo{person}{Rajat~Kumar Pal}, {and} \bibinfo{person}{Deba~Prasad Mandal}.} \bibinfo{year}{2020}\natexlab{}.
\newblock \showarticletitle{Preference enhanced hybrid expertise retrieval system in community question answering services}.
\newblock \bibinfo{journal}{\emph{Decision Support Systems}}  \bibinfo{volume}{129} (\bibinfo{year}{2020}), \bibinfo{pages}{113164}.
\newblock


\bibitem[Lei et~al\mbox{.}(2022)]%
        {lei2022evennet}
\bibfield{author}{\bibinfo{person}{Runlin Lei}, \bibinfo{person}{Zhen Wang}, \bibinfo{person}{Yaliang Li}, \bibinfo{person}{Bolin Ding}, {and} \bibinfo{person}{Zhewei Wei}.} \bibinfo{year}{2022}\natexlab{}.
\newblock \showarticletitle{Evennet: Ignoring odd-hop neighbors improves robustness of graph neural networks}.
\newblock \bibinfo{journal}{\emph{Advances in Neural Information Processing Systems}}  \bibinfo{volume}{35} (\bibinfo{year}{2022}), \bibinfo{pages}{4694--4706}.
\newblock


\bibitem[Lin et~al\mbox{.}(2022)]%
        {lin2022improving}
\bibfield{author}{\bibinfo{person}{Zihan Lin}, \bibinfo{person}{Changxin Tian}, \bibinfo{person}{Yupeng Hou}, {and} \bibinfo{person}{Wayne~Xin Zhao}.} \bibinfo{year}{2022}\natexlab{}.
\newblock \showarticletitle{Improving graph collaborative filtering with neighborhood-enriched contrastive learning}. In \bibinfo{booktitle}{\emph{Proceedings of the ACM Web Conference 2022}}. \bibinfo{pages}{2320--2329}.
\newblock


\bibitem[Liu et~al\mbox{.}(2013)]%
        {liu2013integrating}
\bibfield{author}{\bibinfo{person}{Duen-Ren Liu}, \bibinfo{person}{Yu-Hsuan Chen}, \bibinfo{person}{Wei-Chen Kao}, {and} \bibinfo{person}{Hsiu-Wen Wang}.} \bibinfo{year}{2013}\natexlab{}.
\newblock \showarticletitle{Integrating expert profile, reputation and link analysis for expert finding in question-answering websites}.
\newblock \bibinfo{journal}{\emph{Information processing \& management}} \bibinfo{volume}{49}, \bibinfo{number}{1} (\bibinfo{year}{2013}), \bibinfo{pages}{312--329}.
\newblock


\bibitem[Matsubara et~al\mbox{.}(2020)]%
        {matsubara2020reranking}
\bibfield{author}{\bibinfo{person}{Yoshitomo Matsubara}, \bibinfo{person}{Thuy Vu}, {and} \bibinfo{person}{Alessandro Moschitti}.} \bibinfo{year}{2020}\natexlab{}.
\newblock \showarticletitle{Reranking for efficient transformer-based answer selection}. In \bibinfo{booktitle}{\emph{Proceedings of the 43rd international ACM SIGIR conference on research and development in information retrieval}}. \bibinfo{pages}{1577--1580}.
\newblock


\bibitem[Pei et~al\mbox{.}(2019)]%
        {pei2019personalized}
\bibfield{author}{\bibinfo{person}{Changhua Pei}, \bibinfo{person}{Yi Zhang}, \bibinfo{person}{Yongfeng Zhang}, \bibinfo{person}{Fei Sun}, \bibinfo{person}{Xiao Lin}, \bibinfo{person}{Hanxiao Sun}, \bibinfo{person}{Jian Wu}, \bibinfo{person}{Peng Jiang}, \bibinfo{person}{Junfeng Ge}, \bibinfo{person}{Wenwu Ou}, {et~al\mbox{.}}} \bibinfo{year}{2019}\natexlab{}.
\newblock \showarticletitle{Personalized re-ranking for recommendation}. In \bibinfo{booktitle}{\emph{Proceedings of the 13th ACM conference on recommender systems}}. \bibinfo{pages}{3--11}.
\newblock


\bibitem[Qu et~al\mbox{.}(2009)]%
        {qu2009probabilistic}
\bibfield{author}{\bibinfo{person}{Mingcheng Qu}, \bibinfo{person}{Guang Qiu}, \bibinfo{person}{Xiaofei He}, \bibinfo{person}{Cheng Zhang}, \bibinfo{person}{Hao Wu}, \bibinfo{person}{Jiajun Bu}, {and} \bibinfo{person}{Chun Chen}.} \bibinfo{year}{2009}\natexlab{}.
\newblock \showarticletitle{Probabilistic question recommendation for question answering communities}. In \bibinfo{booktitle}{\emph{Proceedings of the 18th international conference on World wide web}}. \bibinfo{pages}{1229--1230}.
\newblock


\bibitem[Rendle et~al\mbox{.}(2012)]%
        {rendle2012bpr}
\bibfield{author}{\bibinfo{person}{Steffen Rendle}, \bibinfo{person}{Christoph Freudenthaler}, \bibinfo{person}{Zeno Gantner}, {and} \bibinfo{person}{Lars Schmidt-Thieme}.} \bibinfo{year}{2012}\natexlab{}.
\newblock \showarticletitle{BPR: Bayesian personalized ranking from implicit feedback}.
\newblock \bibinfo{journal}{\emph{arXiv preprint arXiv:1205.2618}} (\bibinfo{year}{2012}).
\newblock


\bibitem[Rusch et~al\mbox{.}(2023)]%
        {rusch2023survey}
\bibfield{author}{\bibinfo{person}{T~Konstantin Rusch}, \bibinfo{person}{Michael~M Bronstein}, {and} \bibinfo{person}{Siddhartha Mishra}.} \bibinfo{year}{2023}\natexlab{}.
\newblock \showarticletitle{A survey on oversmoothing in graph neural networks}.
\newblock \bibinfo{journal}{\emph{arXiv preprint arXiv:2303.10993}} (\bibinfo{year}{2023}).
\newblock


\bibitem[Shen et~al\mbox{.}(2009)]%
        {shen2009recommending}
\bibfield{author}{\bibinfo{person}{Jie Shen}, \bibinfo{person}{Wen Shen}, {and} \bibinfo{person}{Xin Fan}.} \bibinfo{year}{2009}\natexlab{}.
\newblock \showarticletitle{Recommending experts in Q\&A communities by weighted HITS algorithm}. In \bibinfo{booktitle}{\emph{2009 international forum on information technology and applications}}, Vol.~\bibinfo{volume}{2}. IEEE, \bibinfo{pages}{151--154}.
\newblock


\bibitem[Tausczik and Pennebaker(2011)]%
        {tausczik2011predicting}
\bibfield{author}{\bibinfo{person}{Yla~R Tausczik} {and} \bibinfo{person}{James~W Pennebaker}.} \bibinfo{year}{2011}\natexlab{}.
\newblock \showarticletitle{Predicting the perceived quality of online mathematics contributions from users' reputations}. In \bibinfo{booktitle}{\emph{Proceedings of the SIGCHI conference on human factors in computing systems}}. \bibinfo{pages}{1885--1888}.
\newblock


\bibitem[Wang et~al\mbox{.}(2019)]%
        {wang2019neural}
\bibfield{author}{\bibinfo{person}{Xiang Wang}, \bibinfo{person}{Xiangnan He}, \bibinfo{person}{Meng Wang}, \bibinfo{person}{Fuli Feng}, {and} \bibinfo{person}{Tat-Seng Chua}.} \bibinfo{year}{2019}\natexlab{}.
\newblock \showarticletitle{Neural graph collaborative filtering}. In \bibinfo{booktitle}{\emph{Proceedings of the 42nd international ACM SIGIR conference on Research and development in Information Retrieval}}. \bibinfo{pages}{165--174}.
\newblock


\bibitem[Wu et~al\mbox{.}(2021)]%
        {wu2021self}
\bibfield{author}{\bibinfo{person}{Jiancan Wu}, \bibinfo{person}{Xiang Wang}, \bibinfo{person}{Fuli Feng}, \bibinfo{person}{Xiangnan He}, \bibinfo{person}{Liang Chen}, \bibinfo{person}{Jianxun Lian}, {and} \bibinfo{person}{Xing Xie}.} \bibinfo{year}{2021}\natexlab{}.
\newblock \showarticletitle{Self-supervised graph learning for recommendation}. In \bibinfo{booktitle}{\emph{Proceedings of the 44th international ACM SIGIR conference on research and development in information retrieval}}. \bibinfo{pages}{726--735}.
\newblock


\bibitem[Wu et~al\mbox{.}(2020)]%
        {wu2020diffnet++}
\bibfield{author}{\bibinfo{person}{Le Wu}, \bibinfo{person}{Junwei Li}, \bibinfo{person}{Peijie Sun}, \bibinfo{person}{Richang Hong}, \bibinfo{person}{Yong Ge}, {and} \bibinfo{person}{Meng Wang}.} \bibinfo{year}{2020}\natexlab{}.
\newblock \showarticletitle{Diffnet++: A neural influence and interest diffusion network for social recommendation}.
\newblock \bibinfo{journal}{\emph{IEEE Transactions on Knowledge and Data Engineering}} \bibinfo{volume}{34}, \bibinfo{number}{10} (\bibinfo{year}{2020}), \bibinfo{pages}{4753--4766}.
\newblock


\bibitem[Wu et~al\mbox{.}(2019)]%
        {wu2019dual}
\bibfield{author}{\bibinfo{person}{Qitian Wu}, \bibinfo{person}{Hengrui Zhang}, \bibinfo{person}{Xiaofeng Gao}, \bibinfo{person}{Peng He}, \bibinfo{person}{Paul Weng}, \bibinfo{person}{Han Gao}, {and} \bibinfo{person}{Guihai Chen}.} \bibinfo{year}{2019}\natexlab{}.
\newblock \showarticletitle{Dual graph attention networks for deep latent representation of multifaceted social effects in recommender systems}. In \bibinfo{booktitle}{\emph{The world wide web conference}}. \bibinfo{pages}{2091--2102}.
\newblock


\bibitem[Yang et~al\mbox{.}(2020)]%
        {yang2020heterogeneous}
\bibfield{author}{\bibinfo{person}{Carl Yang}, \bibinfo{person}{Yuxin Xiao}, \bibinfo{person}{Yu Zhang}, \bibinfo{person}{Yizhou Sun}, {and} \bibinfo{person}{Jiawei Han}.} \bibinfo{year}{2020}\natexlab{}.
\newblock \showarticletitle{Heterogeneous network representation learning: A unified framework with survey and benchmark}.
\newblock \bibinfo{journal}{\emph{IEEE Transactions on Knowledge and Data Engineering}} \bibinfo{volume}{34}, \bibinfo{number}{10} (\bibinfo{year}{2020}), \bibinfo{pages}{4854--4873}.
\newblock


\bibitem[Ye et~al\mbox{.}(2023)]%
        {ye2023towards}
\bibfield{author}{\bibinfo{person}{Haibo Ye}, \bibinfo{person}{Xinjie Li}, \bibinfo{person}{Yuan Yao}, {and} \bibinfo{person}{Hanghang Tong}.} \bibinfo{year}{2023}\natexlab{}.
\newblock \showarticletitle{Towards robust neural graph collaborative filtering via structure denoising and embedding perturbation}.
\newblock \bibinfo{journal}{\emph{ACM Transactions on Information Systems}} \bibinfo{volume}{41}, \bibinfo{number}{3} (\bibinfo{year}{2023}), \bibinfo{pages}{1--28}.
\newblock


\bibitem[Yu et~al\mbox{.}(2023)]%
        {yu2023xsimgcl}
\bibfield{author}{\bibinfo{person}{Junliang Yu}, \bibinfo{person}{Xin Xia}, \bibinfo{person}{Tong Chen}, \bibinfo{person}{Lizhen Cui}, \bibinfo{person}{Nguyen Quoc~Viet Hung}, {and} \bibinfo{person}{Hongzhi Yin}.} \bibinfo{year}{2023}\natexlab{}.
\newblock \showarticletitle{XSimGCL: Towards extremely simple graph contrastive learning for recommendation}.
\newblock \bibinfo{journal}{\emph{IEEE Transactions on Knowledge and Data Engineering}} (\bibinfo{year}{2023}).
\newblock


\bibitem[Yu et~al\mbox{.}(2021)]%
        {yu2021self}
\bibfield{author}{\bibinfo{person}{Junliang Yu}, \bibinfo{person}{Hongzhi Yin}, \bibinfo{person}{Jundong Li}, \bibinfo{person}{Qinyong Wang}, \bibinfo{person}{Nguyen Quoc~Viet Hung}, {and} \bibinfo{person}{Xiangliang Zhang}.} \bibinfo{year}{2021}\natexlab{}.
\newblock \showarticletitle{Self-supervised multi-channel hypergraph convolutional network for social recommendation}. In \bibinfo{booktitle}{\emph{Proceedings of the web conference 2021}}. \bibinfo{pages}{413--424}.
\newblock


\bibitem[Yu et~al\mbox{.}(2022)]%
        {yu2022graph}
\bibfield{author}{\bibinfo{person}{Junliang Yu}, \bibinfo{person}{Hongzhi Yin}, \bibinfo{person}{Xin Xia}, \bibinfo{person}{Tong Chen}, \bibinfo{person}{Lizhen Cui}, {and} \bibinfo{person}{Quoc Viet~Hung Nguyen}.} \bibinfo{year}{2022}\natexlab{}.
\newblock \showarticletitle{Are graph augmentations necessary? simple graph contrastive learning for recommendation}. In \bibinfo{booktitle}{\emph{Proceedings of the 45th international ACM SIGIR conference on research and development in information retrieval}}. \bibinfo{pages}{1294--1303}.
\newblock


\bibitem[Zhang et~al\mbox{.}(2007)]%
        {zhang2007expertise}
\bibfield{author}{\bibinfo{person}{Jun Zhang}, \bibinfo{person}{Mark~S Ackerman}, {and} \bibinfo{person}{Lada Adamic}.} \bibinfo{year}{2007}\natexlab{}.
\newblock \showarticletitle{Expertise networks in online communities: structure and algorithms}. In \bibinfo{booktitle}{\emph{Proceedings of the 16th international conference on World Wide Web}}. \bibinfo{pages}{221--230}.
\newblock


\bibitem[Zhang et~al\mbox{.}(2020)]%
        {zhang2020answer}
\bibfield{author}{\bibinfo{person}{Wenxuan Zhang}, \bibinfo{person}{Yang Deng}, {and} \bibinfo{person}{Wai Lam}.} \bibinfo{year}{2020}\natexlab{}.
\newblock \showarticletitle{Answer ranking for product-related questions via multiple semantic relations modeling}. In \bibinfo{booktitle}{\emph{Proceedings of the 43rd International ACM SIGIR conference on research and development in Information Retrieval}}. \bibinfo{pages}{569--578}.
\newblock


\bibitem[Zhang et~al\mbox{.}(2022)]%
        {zhang2022counteracting}
\bibfield{author}{\bibinfo{person}{Xiao Zhang}, \bibinfo{person}{Sunhao Dai}, \bibinfo{person}{Jun Xu}, \bibinfo{person}{Zhenhua Dong}, \bibinfo{person}{Quanyu Dai}, {and} \bibinfo{person}{Ji-Rong Wen}.} \bibinfo{year}{2022}\natexlab{}.
\newblock \showarticletitle{Counteracting user attention bias in music streaming recommendation via reward modification}. In \bibinfo{booktitle}{\emph{Proceedings of the 28th ACM SIGKDD Conference on Knowledge Discovery and Data Mining}}. \bibinfo{pages}{2504--2514}.
\newblock


\bibitem[Zhao et~al\mbox{.}(2020)]%
        {zhao2020revisiting}
\bibfield{author}{\bibinfo{person}{Wayne~Xin Zhao}, \bibinfo{person}{Junhua Chen}, \bibinfo{person}{Pengfei Wang}, \bibinfo{person}{Qi Gu}, {and} \bibinfo{person}{Ji-Rong Wen}.} \bibinfo{year}{2020}\natexlab{}.
\newblock \showarticletitle{Revisiting alternative experimental settings for evaluating top-n item recommendation algorithms}. In \bibinfo{booktitle}{\emph{Proceedings of the 29th ACM International Conference on Information \& Knowledge Management}}. \bibinfo{pages}{2329--2332}.
\newblock


\bibitem[Zhao et~al\mbox{.}(2022)]%
        {zhao2022recbole2}
\bibfield{author}{\bibinfo{person}{Wayne~Xin Zhao}, \bibinfo{person}{Yupeng Hou}, \bibinfo{person}{Xingyu Pan}, \bibinfo{person}{Chen Yang}, \bibinfo{person}{Zeyu Zhang}, \bibinfo{person}{Zihan Lin}, \bibinfo{person}{Jingsen Zhang}, \bibinfo{person}{Shuqing Bian}, \bibinfo{person}{Jiakai Tang}, \bibinfo{person}{Wenqi Sun}, \bibinfo{person}{Yushuo Chen}, \bibinfo{person}{Lanling Xu}, \bibinfo{person}{Gaowei Zhang}, \bibinfo{person}{Zhen Tian}, \bibinfo{person}{Changxin Tian}, \bibinfo{person}{Shanlei Mu}, \bibinfo{person}{Xinyan Fan}, \bibinfo{person}{Xu Chen}, {and} \bibinfo{person}{Ji-Rong Wen}.} \bibinfo{year}{2022}\natexlab{}.
\newblock \showarticletitle{RecBole 2.0: Towards a More Up-to-Date Recommendation Library}. In \bibinfo{booktitle}{\emph{{CIKM}}}.
\newblock


\bibitem[Zhao et~al\mbox{.}(2021)]%
        {zhao2021recbole}
\bibfield{author}{\bibinfo{person}{Wayne~Xin Zhao}, \bibinfo{person}{Shanlei Mu}, \bibinfo{person}{Yupeng Hou}, \bibinfo{person}{Zihan Lin}, \bibinfo{person}{Yushuo Chen}, \bibinfo{person}{Xingyu Pan}, \bibinfo{person}{Kaiyuan Li}, \bibinfo{person}{Yujie Lu}, \bibinfo{person}{Hui Wang}, \bibinfo{person}{Changxin Tian}, \bibinfo{person}{Yingqian Min}, \bibinfo{person}{Zhichao Feng}, \bibinfo{person}{Xinyan Fan}, \bibinfo{person}{Xu Chen}, \bibinfo{person}{Pengfei Wang}, \bibinfo{person}{Wendi Ji}, \bibinfo{person}{Yaliang Li}, \bibinfo{person}{Xiaoling Wang}, {and} \bibinfo{person}{Ji{-}Rong Wen}.} \bibinfo{year}{2021}\natexlab{}.
\newblock \showarticletitle{RecBole: Towards a Unified, Comprehensive and Efficient Framework for Recommendation Algorithms}. In \bibinfo{booktitle}{\emph{{CIKM}}}. \bibinfo{publisher}{{ACM}}, \bibinfo{pages}{4653--4664}.
\newblock


\bibitem[Zhao et~al\mbox{.}(2016)]%
        {zhao2016expert}
\bibfield{author}{\bibinfo{person}{Zhou Zhao}, \bibinfo{person}{Qifan Yang}, \bibinfo{person}{Deng Cai}, \bibinfo{person}{Xiaofei He}, {and} \bibinfo{person}{Yueting Zhuang}.} \bibinfo{year}{2016}\natexlab{}.
\newblock \showarticletitle{Expert finding for community-based question answering via ranking metric network learning.}. In \bibinfo{booktitle}{\emph{Ijcai}}, Vol.~\bibinfo{volume}{16}. \bibinfo{pages}{3000--3006}.
\newblock


\bibitem[Zhao et~al\mbox{.}(2014)]%
        {zhao2014expert}
\bibfield{author}{\bibinfo{person}{Zhou Zhao}, \bibinfo{person}{Lijun Zhang}, \bibinfo{person}{Xiaofei He}, {and} \bibinfo{person}{Wilfred Ng}.} \bibinfo{year}{2014}\natexlab{}.
\newblock \showarticletitle{Expert finding for question answering via graph regularized matrix completion}.
\newblock \bibinfo{journal}{\emph{IEEE Transactions on Knowledge and Data Engineering}} \bibinfo{volume}{27}, \bibinfo{number}{4} (\bibinfo{year}{2014}), \bibinfo{pages}{993--1004}.
\newblock


\end{thebibliography}

\end{document}